\documentclass[preprint,12pt,fleqn]{elsarticle}

\usepackage{graphicx}
\usepackage{amssymb}
\usepackage{amsthm}
\usepackage{amsmath}
\usepackage{algorithm}
\usepackage{algpseudocode}
\usepackage{hyperref}
\usepackage{color}
\usepackage{lineno}
\usepackage[colorinlistoftodos,textsize=small]{todonotes}

\renewcommand{\Im}{\mathrm{Im}}

\journal{Computer Physics Communications}

\begin{document}

\begin{frontmatter}

\title{ELSI: A Unified Software Interface for Kohn-Sham Electronic Structure Solvers}


\author[duke_mems]{Victor Wen-zhe Yu}
\author[imperial]{Fabiano Corsetti}
\author[icmab]{Alberto Garc\'{i}a}
\author[duke_mems]{William P. Huhn}
\author[berkeley]{Mathias Jacquelin}
\author[berkeley,uc_berkeley]{Weile Jia}
\author[duke_mems]{Bj\"{o}rn Lange}
\author[berkeley,uc_berkeley]{Lin Lin}
\author[duke_math]{Jianfeng Lu}
\author[duke_mems]{Wenhui Mi}
\author[duke_mems]{Ali Seifitokaldani}
\author[argonne]{\'{A}lvaro V\'{a}zquez-Mayagoitia}
\author[berkeley]{Chao Yang}
\author[duke_math]{Haizhao Yang}
\author[duke_mems]{Volker Blum  \corref{ca}}
\ead{volker.blum@duke.edu}
\cortext[ca]{Corresponding author.}

\address[duke_mems]{Department of Mechanical Engineering and Materials Science, Duke University, Durham, NC 27707}
\address[imperial]{Departments of Materials and Physics, and the Thomas Young Centre for Theory and Simulation of Materials, Imperial College London, London SW7 2AZ, United Kingdom}
\address[icmab]{Institut de Ci\`{e}ncia de Materials de Barcelona (ICMAB-CSIC), Bellaterra E-08193, Spain}
\address[berkeley]{Computational Research Division, Lawrence Berkeley National Laboratory, Berkeley, CA 94720}
\address[uc_berkeley]{Department of Mathematics, University of California, Berkeley, CA 94720}
\address[duke_math]{Department of Mathematics, Duke University, Durham, NC 27707}
\address[argonne]{Argonne Leadership Computing Facility, Argonne National Laboratory, Argonne, IL 60439}

\begin{abstract}
Solving the electronic structure from a generalized or standard eigenproblem is often the bottleneck in large scale calculations based on Kohn-Sham density-functional theory. This problem must be addressed by essentially all current electronic structure codes, based on similar matrix expressions, and by high-performance computation. We here present a unified software interface, ELSI, to access different strategies that address the Kohn-Sham eigenvalue problem. Currently supported algorithms include the dense generalized eigensolver library ELPA, the orbital minimization method implemented in libOMM, and the pole expansion and selected inversion (PEXSI) approach with lower computational complexity for semilocal density functionals. The ELSI interface aims to simplify the implementation and optimal use of the different strategies, by offering (a) a unified software framework designed for the electronic structure solvers in Kohn-Sham density-functional theory; (b) reasonable default parameters for a chosen solver; (c) automatic conversion between input and internal working matrix formats, and in the future (d) recommendation of the optimal solver depending on the specific problem. Comparative benchmarks are shown for system sizes up to 11,520 atoms (172,800 basis functions) on distributed memory supercomputing architectures.
\end{abstract}


\end{frontmatter}

{\bf \noindent PROGRAM SUMMARY}\\
\begin{small}
\noindent
{\em Program title: ELSI Interface}\\
{\em Licensing provisions: BSD 3-clause}\\
{\em Distribution format: tar.gz}\\
{\em Programming language: Fortran 2003, with interface to C/C++}\\
{\em External routines/libraries: MPI, BLAS, LAPACK, ScaLAPACK, ELPA, libOMM, PEXSI, ParMETIS, SuperLU\_DIST}\\
{\em Operating system: Unix-like (Linux, macOS), Windows (not tested)}\\
{\em Nature of problem: Solving the electronic structure from a generalized or standard eigenvalue problem in calculations based on Kohn-Sham density functional theory (KS-DFT).}\\
{\em Solution method: To connect the KS-DFT codes and the KS electronic structure solvers, ELSI provides a unified software interface with reasonable default parameters, hierarchical control over the interface and the solvers, and automatic conversions between input and internal working matrix formats. Supported solvers are: ELPA (dense generalized eigensolver), libOMM (orbital minimization method), and PEXSI (pole expansion and selected inversion method).}\\
{\em Restrictions: The ELSI interface requires complete information of the Hamiltonian matrix.}\\
\end{small}


\section{Introduction}
\label{sec:intro}
Molecular and materials simulations based on Kohn-Sham (KS) \cite{ks_kohn_1965} and generalized Kohn-Sham (gKS) \cite{gks_becke_1993,gks_seidl_1996} density-functional theory (DFT) are widely used to provide atomic-scale insights, understanding, and predictions across a wide range of disciplines in the sciences and in engineering. The number of DFT-related publications has grown rapidly over recent decades \cite{dft_ghiringhelli_2016,dft_haunschild_2016,dft_mavropoulos_2017}, exceeding 20,000 in 2016 \cite{dft_mavropoulos_2017}. In particular, simulations based on semilocal and hybrid density functionals serve as the production workhorses for a broad range of applications. Advances in both computational methods and high-performance computing hardware render it feasible to model large systems consisting of thousands of atoms, and linear scaling KS-DFT \cite{on_goedecker_1999,on_bowler_2012,onetep_skylaris_2005} can reach system sizes of millions of atoms \cite{million_bowler_2010,million_vandevondele_2012}. Higher levels of density functional approximations, like the Random Phase Approximation (RPA), can be formulated to scale linearly with system size as well \cite{on_schurkus_2016,on_luenser_2017}.

However, approaches for which the computational effort scales lower than $O(N^3)$, where $N$ is some measure of the system size, are, arguably, not yet fully established as mainstream methods of the field. There are several reasons for this status. Perhaps the simplest reason is that formally $O(N^3)$ scaling approaches by solving an algebraic eigenvalue problem are generally applicable to any class of system, and the computational effort associated with them has a low prefactor, i.e., they are advantageous to use for systems comprised of up to roughly a few thousands of atoms, which account for the bulk of KS-DFT applications. In contrast, the transition to lower-scaling solution methods for larger systems is not necessarily simple. Such alternatives are typically restricted to certain classes of systems or problems. The transition is, therefore, not trivial to automate, requiring specific intervention and sometimes specialist knowledge by its users. This creates hurdles both from a user point of view (complexity of choice) and from a developer point of view (replication of often complex infrastructure to implement a particular method efficiently). The KS eigenvalue problem is thus in practice a bottleneck of KS-DFT simulations on current HPC architectures and for system sizes significantly exceeding several thousands of atoms.

We here present a software infrastructure, ELSI, that simplifies the approach to overcome the Kohn-Sham eigenproblem bottleneck as much as possible for electronic structure users and developers. ELSI provides an integrated and extendable interface to multiple strategies targeting the KS eigenproblem (referred to as Kohn-Sham electronic structure solvers throughout this paper). It presently (version: 2017.05) supports three solvers: ELPA (Eigenvalue soLvers for Petaflop-Applications) \cite{elpa_auckenthaler_2011,elpa_marek_2014}, libOMM (Orbital Minimization Method) \cite{libomm_corsetti_2014}, and PEXSI (Pole EXpansion and Selected Inversion) \cite{pexsi_lin_2009,pexsi_lin_2013,pexsi_lin_2014}. For the future, ELSI is expressly intended to integrate further solvers such as the linear-scaling solver CheSS(CHEbyshev Sparse Solvers) \cite{chess_mohr_2017}, the iterative solver SIPs (Shift-and-Invert Parallel Spectral transformation eigensolver) \cite{sips_keceli_2016}, and others. By design, ELSI is an open infrastructure, intended to serve a community, and it can and should be flexibly adaptable to new solvers and new electronic structure codes' needs in the future. In this paper, we describe the outline and basic principles of ELSI, as well as a comparative assessment of the three solution strategies that are already supported in ELSI as of its 2017.05 release. The software presented here is a structural foundation that is already working in several electronic structure codes, and we expect it to become a focal point for new developments and solver cross-comparisons in the future.

\section{Kohn-Sham Density-Functional Theory}
\label{sec:ksdft}
In KS-DFT \cite{ks_kohn_1965}, the many-electron problem for the Born-Oppenheimer electronic ground state is reduced to a system of single particle equations known as the Kohn-Sham equations

\begin{equation}
\label{eq:ks}
\hat{h}^\text{KS} \psi_l = \epsilon_l \psi_l ,
\end{equation}

\noindent where $\psi_l$ and $\epsilon_l$ are Kohn-Sham orbitals and their associated eigenenergies, and $\hat{h}^\text{KS}$ denotes the Kohn-Sham Hamiltonian:

\begin{equation}
\label{eq:ham}
\hat{h}^\text{KS} = \hat{t}_\text{s} + \hat{v}_\text{es} + \hat{v}_\text{xc} + \hat{v}_\text{ext} ,
\end{equation}

\noindent which includes the kinetic energy $\hat{t}_\text{s}$, the average electrostatic potential of the electron density and of the nuclei $\hat{v}_\text{es}$ (i.e. the Hartree potential), the exchange-correlation potential $\hat{v}_\text{xc}$, and possible additional potential terms $\hat{v}_\text{ext}$ from external electromagnetic fields.

In almost all practical approaches, $N_\text{basis}$ basis functions ${\phi_i(\boldsymbol{r})}$ are employed to approximately expand the Kohn-Sham orbitals:

\begin{equation}
\label{eq:basis_expansion}
\psi_l(\boldsymbol{r}) = \sum_{j=1}^{N_\text{basis}} c_{jl} \phi_j(\boldsymbol{r}) .
\end{equation}

\noindent The choice of basis set is one of the critical decisions in the design of an electronic structure code \cite{elephant_jensen_2017}. Using non-orthogonal basis functions (e.g., Gaussian functions \cite{elephant_jensen_2017,gto_szabo_1989,cp2k_hutter_2014,gaussian_frisch_2016,nwchem_valiev_2010,qchem_shao_2015,turbomole_furche_2014}, Slater functions \cite{sto_slater_1930,adf_velde_2001}, numeric atom-centered orbitals \cite{atk_web,dmol3_delley_1990,fhiaims_blum_2009,fplo_koepernik_1999,openmx_ozaki_2003,siesta_soler_2002}, (linearized) augmented plane waves \cite{lapw_singh_1994,elk_web,exciting_gulans_2014,fleur_blugel_1987,wien2k_blaha_2001}, finite elements \cite{fe_pask_2001}) in Eq. \ref{eq:basis_expansion} converts Eq. \ref{eq:ks} to a generalized eigenvalue problem

\begin{equation}
\label{eq:generalized_evp}
\sum_j h_{ij} c_{jl} = \epsilon_l \sum_j s_{ij} c_{jl} ,
\end{equation}

\noindent where $h_{ij}$ and $s_{ij}$ are the elements of the Hamiltonian matrix $\boldsymbol{H}$ and the overlap matrix $\boldsymbol{S}$, which can be computed through numerical integrations:

\begin{equation}
\label{eq:ham_ovlp_integration}
\begin{split}
h_{ij} & = \int d^3 r [\phi_i^*(\boldsymbol{r}) \hat{h}^\text{KS} \phi_j(\boldsymbol{r})] ,\\
s_{ij} & = \int d^3 r [\phi_i^*(\boldsymbol{r}) \phi_j(\boldsymbol{r})] .
\end{split}
\end{equation}

Eq. \ref{eq:generalized_evp} can thus be expressed in the following matrix form:

\begin{equation}
\label{eq:generalized_evp_matrix}
\boldsymbol{H} \boldsymbol{c} = \boldsymbol{\epsilon} \boldsymbol{S} \boldsymbol{c} .
\end{equation}

\noindent Here, the matrix $\boldsymbol{c}$ and diagonal matrix $\boldsymbol{\epsilon}$ contain the eigenvectors and eigenvalues, respectively, of the eigensystem of the matrices $\boldsymbol{H}$ and $\boldsymbol{S}$.

When using orthonormal basis sets (e.g., plane waves \cite{onetep_skylaris_2005,planewave_martin_2004,abinit_gonze_2009,castep_clark_2005,qe_giannozzi_2009,vasp_kresse_1996}, multi-resolution wavelets \cite{bigdft_mohr_2015,madness_harrison_2016,mrchem_web}, adaptive local basis set \cite{alb_lin_2012,dgdft_hu_2015}, grid-discretization based approaches \cite{rmg_hernholc_2008,parsec_kronik_2006}), the eigenproblem described in Eq. \ref{eq:generalized_evp_matrix} reduces to a standard form where $s_{ij}=\delta_{ij}$, or even can be circumvented completely by solving the KS equations in an integral formulation \cite{elephant_jensen_2017}.

The explicit solution of Eq. \ref{eq:generalized_evp} or \ref{eq:generalized_evp_matrix} yields the Kohn-Sham orbitals $\psi_i$, from which the electron density $n(\boldsymbol{r})$ can be computed following an orbital-based method that scales as $O(N^2)$:

\begin{equation}
\label{eq:orbital_update}
n(\boldsymbol{r}) = \sum_{j=1}^{N_\text{basis}} f_l \psi_l^*(\boldsymbol{r}) \psi_l(\boldsymbol{r}) ,
\end{equation}

\noindent where $f_l$ denotes the occupation number of each orbital. In an actual computation, it is sufficient to perform the summation only for the occupied ($f_l > 0$) orbitals. The ratio of occupied orbitals to the total number of basis functions can be below 1\% for plane wave basis sets, whereas with some localized basis sets, fewer basis functions are required, leading to a larger fraction of occupied states typically between 10\% and 40\%.

An alternative method that scales as $O(N)$ can be employed for localized basis functions:

\begin{equation}
\label{eq:density_matrix_update}
n(\boldsymbol{r}) = \sum_{i,j}^{N_\text{basis}} \phi_i^*(\boldsymbol{r}) n_{ij} \phi_j(\boldsymbol{r}) ,
\end{equation}

\noindent with $n_{ij}$ being the elements of the density matrix that need to be computed before the density update:

\begin{equation}
\label{eq:density_matrix}
n_{ij} = \sum_{l=1}^{N_\text{basis}} f_l c_{il} c_{jl} .
\end{equation}

Due to the dependence of $\boldsymbol{H}$ on $\psi_l$ via the density and the potentials, Eqs. \ref{eq:generalized_evp} and \ref{eq:generalized_evp_matrix} are in fact non-linear eigenvalue problems, and therefore must be solved in an iterative fashion. The most commonly used method is the self-consistent field (SCF) or fixed-point iteration approach. To achieve self-consistency, the electron density needs to be updated in every iteration until converged to an acceptable level. From a viewpoint of computational complexity, almost all standard pieces of solving the Kohn-Sham equations can be formulated in a linear scaling fashion with respect to the system size. The only piece that can not, in all cases and for all semilocal and hybrid functionals, be easily addressed in an $O(N)$ fashion is the Kohn-Sham eigenvalue problem described in Eq. \ref{eq:generalized_evp}.

\section{Kohn-Sham Electronic Structure Solvers Supported by ELSI}
\label{sec:solvers}
\subsection{ELPA: Eigenvalue soLvers for Petaflop-Applications}
\label{subsec:elpa_intro}
The Kohn-Sham eigenvalue problem in Eq. \ref{eq:generalized_evp} can be explicitly solved by traditional (tri)diagonalization \cite{evp_golub_2013}. In ELSI, the massively parallel direct solver ELPA \cite{elpa_auckenthaler_2011,elpa_marek_2014} facilitates the solution of symmetric or Hermitian eigenproblems on high-performance computers. It was initially designed for distributed memory architectures, then extended to exploit multi-threading parallelism, and is subject to ongoing work for GPU acceleration.

In ELPA, the generalized eigenproblem in Eq. \ref{eq:generalized_evp_matrix} is first transformed to the standard form by Cholesky decomposition of the overlap matrix $\boldsymbol{S}$:

\begin{equation}
\label{eq:cholesky}
\boldsymbol{S} = \boldsymbol{L} \boldsymbol{L}^* ,
\end{equation}

\noindent where $\boldsymbol{L}$ is a lower triangular matrix. Eq. \ref{eq:generalized_evp_matrix} is then transformed by applying the Cholesky factor:

\begin{equation}
\label{eq:after_cholesky}
\boldsymbol{\tilde{H}} \boldsymbol{\tilde{c}} = \boldsymbol{\epsilon} \boldsymbol{\tilde{c}}
\end{equation}

\noindent with $\boldsymbol{\tilde{H}} = \boldsymbol{L}^{-1} \boldsymbol{H} (\boldsymbol{L}^*)^{-1}$ and $\boldsymbol{\tilde{c}} = \boldsymbol{L}^* \boldsymbol{c}$.

Then, the standard eigenproblem is either directly reduced to the tridiagonal form

\begin{equation}
\label{eq:elpa1}
\boldsymbol{T} = \boldsymbol{Q} \boldsymbol{\tilde{H}} \boldsymbol{Q}^* ,
\end{equation}

\noindent or first reduced to a banded intermediate form, then to the tridiagonal form \cite{sbr_bischof_2000}:

\begin{equation}
\label{eq:elpa2}
\begin{split}
\boldsymbol{B} & = \boldsymbol{Q}_1 \boldsymbol{\tilde{H}} \boldsymbol{Q}_1^* ,\\
\boldsymbol{T} & = \boldsymbol{Q}_2 \boldsymbol{B} \boldsymbol{Q}_2^* .
\end{split}
\end{equation}

\noindent In Eqs. \ref{eq:elpa1} and \ref{eq:elpa2}, $\boldsymbol{Q}$, $\boldsymbol{Q}_1$, $\boldsymbol{Q}_2$ are transformation matrices; $\boldsymbol{T}$ is a tridiagonal matrix; $\boldsymbol{B}$ is a banded matrix.

\begin{figure*}[h!]
\centering
\includegraphics[width=0.9\textwidth]{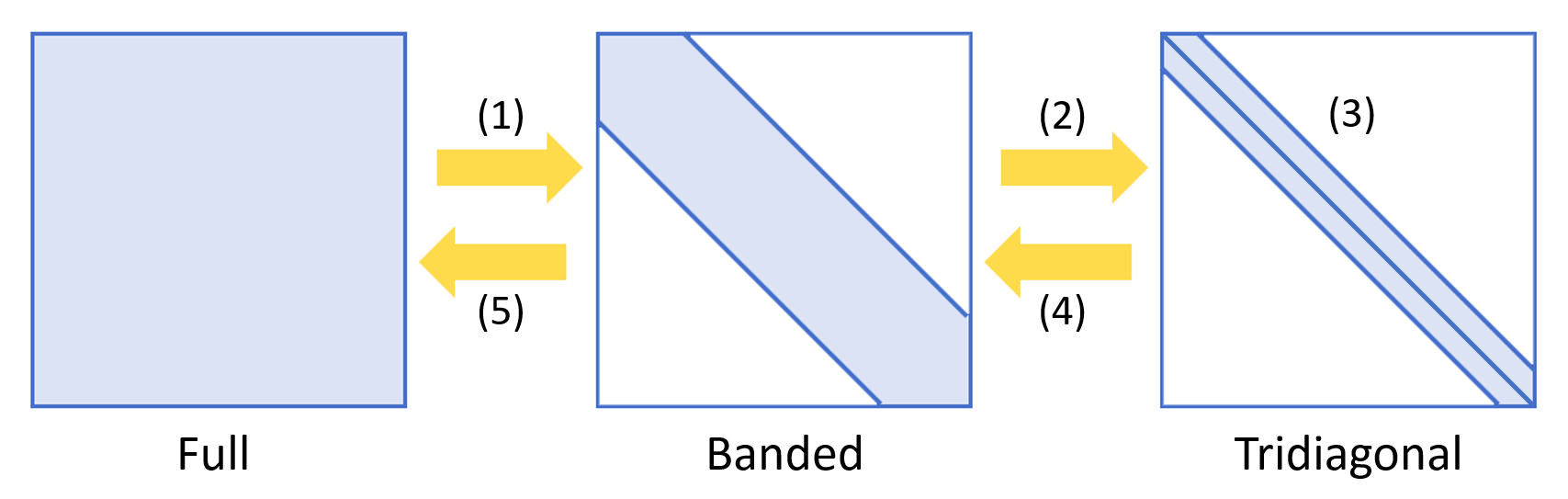}
\caption{Five computational steps of the ELPA eigensolver with two-stage tridiagonalization. (1) Reduction of the full matrix to a banded form. (2) Reduction of the banded matrix to a tridiagonal form. (3) Solution of the eigenvalues and eigenvectors of the tridiagonal system. (4) Back-transformation of the eigenvectors to the banded form. (5) Back-transformation of the eigenvectors to the original full form. This figure is redesigned based on Fig. 1 in Ref. \cite{elpa_marek_2014}.}
\label{fig:elpa2}
\end{figure*}

The key steps of the two-stage tridiagonalization algorithm implemented in ELPA are reviewed in Fig. \ref{fig:elpa2}. Steps (1) and (2) correspond to Eq. \ref{eq:elpa2}, i.e. the transformations to the banded and tridiagonal forms. Step (3) corresponds to the solution of the actual eigenvalue problem by a divide-and-conquer approach \cite{elpa_auckenthaler_2011,dc_cuppen_1980}, which can be restricted to compute only a fraction of the eigenvectors. Finally, the computed eigenvectors are transformed back into the representations corresponding to the banded (step (4)) and standard forms (step (5)) of the problem. Compared to the one-step tridiagonalization (Eq. \ref{eq:elpa1}), the two-step algorithm introduces two additional steps (steps (1) and (5) in Fig. \ref{fig:elpa2}). Still, the two-step approach has been shown to enable faster computation and better parallel scalability than the one-step approach on present-day computers \cite{elpa_marek_2014}. Specifically, the matrix-vector operations (BLAS level-2 routines) in the one-step tridiagonalization can be mostly replaced by more efficient matrix-matrix operations (BLAS level-3 routines) in the two-step version of the algorithm \cite{2step_bischof_1994}. Since steps 2 and 4 pertain to forward and back transformations between banded and tridiagonal matrices only, the resulting transformations can be efficiently grouped to minimize computational overhead, especially for the back transformation in step (4) \cite{elpa_auckenthaler_2011}. The computational workload associated with step (4) is further alleviated in KS-DFT calculations if only a small fraction of the eigenvectors representing the lowest eigenstates is required, and by architecture-specific linear-algebra ``kernels'' provided with the ELPA library \cite{elpa_auckenthaler_2011,elpa_marek_2014}.

Since ELPA employs the same 2D block-cyclic matrix distribution as does the ScaLAPACK library \cite{scalapack_blackford_1997} (by way of the basic linear algebra communication subroutines (BLACS) \cite{blacs_anderson_1991}), it can easily be substituted into existing codes that already support parallel linear algebra by ScaLAPACK.

\subsection{libOMM: Orbital Minimization Method}
\label{subsec:omm_intro}
Instead of diagonalizing the $N_\text{basis} \times N_\text{basis}$ eigenproblem, the orbital minimization method (OMM) relies on efficient iterative algorithms to directly minimize an unconstrained energy functional using a set of auxiliary orbitals that are not the Kohn-Sham orbitals $\phi_i$. These auxiliary orbitals are then used to obtain the density matrix of the system. Specifically, the OMM employs $N_\text{W} = N_\text{electron}/2$ non-orthogonal Wannier functions $\chi_k$ to represent the occupied subspace of a system with $N_\text{electron}$ electrons:

\begin{equation}
\label{eq:wannier}
\chi_k = \sum_{j=1}^{N_\text{basis}} W_{kj} \phi_j .
\end{equation}

\begin{figure*}[h!]
\centering
\includegraphics[width=0.9\textwidth]{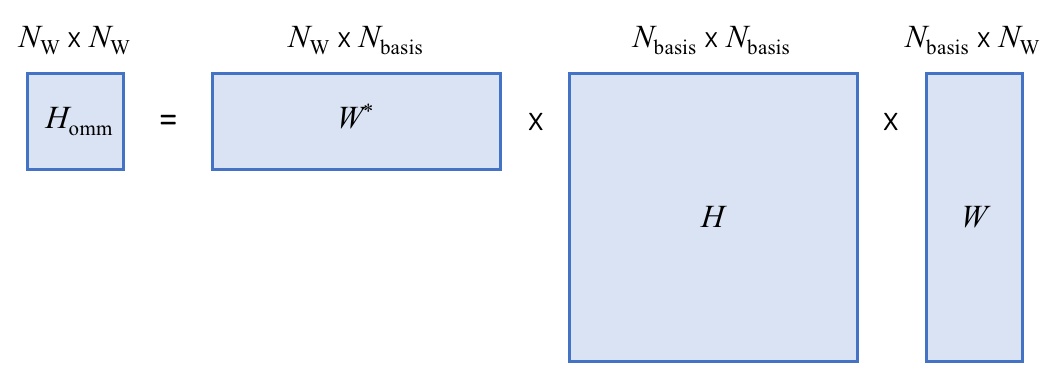}
\caption{Schematic representation of sizes of Hamiltonian matrix before and after applying the Wannier function transformation in the orbital minimization method. Matrix dimensions are shown above the matrices. $N_\text{W}$: Number of Wannier functions. $N_\text{basis}$: Number of basis functions.}
\label{fig:omm}
\end{figure*}

\noindent For non-spinpolarized systems, the index $k$ runs from $1$ to $N_\text{W}$. Then the matrices $\boldsymbol{H}$ and $\boldsymbol{S}$ in the occupied subspace become

\begin{equation}
\label{eq:reduced_ham_ovlp}
\begin{split}
\boldsymbol{H_\text{omm}} & = \boldsymbol{W}^* \boldsymbol{H} \boldsymbol{W} ,\\
\boldsymbol{S_\text{omm}} & = \boldsymbol{W}^* \boldsymbol{S} \boldsymbol{W} ,
\end{split}
\end{equation}

\noindent where $\boldsymbol{W}$ is the coefficient matrix of the Wannier functions. The size change of the Hamiltonian matrix facilitated by Eq. \ref{eq:reduced_ham_ovlp} is illustrated in Fig. \ref{fig:omm}.

The OMM energy functional is defined as

\begin{equation}
\label{eq:omm_energy}
E[\boldsymbol{W}] = 4 Tr[\boldsymbol{H_\text{omm}}] - 2 Tr[\boldsymbol{S_\text{omm} H_\text{omm}}] .
\end{equation}

\noindent This functional, when minimized with respect to the coefficients of Wannier functions $\boldsymbol{W}$, can be shown to be equal to the sum of the lowest $N_\text{electron}/2$ eigenvalues of the original KS eigenproblem \cite{omm_mauri_1993,omm_ordejon_1993,omm_mauri_1994,omm_ordejon_1995}. Furthermore, the Wannier functions are driven towards perfect orthonormality at this minimum. The density matrix is then constructed from the $\boldsymbol{W}$ that minimizes $E[\boldsymbol{W}]$. Although this density matrix is sufficient for the electron density update following Eq. \ref{eq:density_matrix_update}, compared to the density matrix in Eq. \ref{eq:density_matrix}, it is obvious that the occupation numbers are restricted to be integers (1 for occupied; 0 for unoccupied) in this method. Without knowledge of individual eigenstates, the OMM cannot handle systems with fractional occupation numbers resulting, e.g., from a finite electronic temperature, such as is typically required for metals.

Compared to other minimization methods with the orthonormality constraint of eigenstates \cite{vasp_kresse_1996,min_teter_1989,min_payne_1992}, the advantage of the OMM is that it only requires an unconstrained minimization without an explicit orthonormalization step. This makes the OMM a good candidate for linear scaling DFT; indeed, the method was originally developed in this context \cite{omm_mauri_1993,omm_ordejon_1993,omm_mauri_1994,omm_ordejon_1995}. However, in order to do so, it is necessary to spatially confine the Wannier functions by imposing a certain sparsity to $\boldsymbol{W}$. This introduces a number of technical difficulties which have ultimately required the development of more involved algorithms \cite{omm_mauri_1993,omm_mauri_1994,omm_kim_1995}. The properties of the original OMM functional with unconstrained Wannier functions have nevertheless been found to result in an extremely efficient iterative solver with conventional cubic scaling but a smaller prefactor than diagonalization. This approach has been taken by the new implementation in libOMM \cite{libomm_corsetti_2014}. It should be noted that for finite-range basis sets in which $\boldsymbol{W}$ is formally sparse, this sparsity can be taken into account to reduce the scaling of the matrix-matrix product $\boldsymbol{HW}$ from cubic to quadratic, thus effectively eliminating the most expensive matrix operation in the algorithm. The minimization of the OMM energy functional in Eq. \ref{eq:omm_energy} is carried out in libOMM by using the conjugate-gradient (CG) method with an efficient preconditioning using the kinetic energy matrix, as described in Ref. \cite{libomm_corsetti_2014}.

\subsection{PEXSI: Pole EXpansion and Selected Inversion}
\label{subsec:pexsi_intro}
The density matrix in Eq. \ref{eq:density_matrix} is associated with the Kohn-Sham orbitals and their occupation numbers $f_l$, which are given by the Fermi-Dirac distribution function \cite{fermi_mermin_1964}:

\begin{equation}
\label{eq:fermi_dirac}
f_l = \frac{1}{1+e^{\frac{\epsilon_l-\mu}{k_B T}}}.
\end{equation}

\noindent Here $k_B$ is the Boltzmann constant, $T$ is the temperature, and $\mu$ is the chemical potential that is determined by the normalization condition

\begin{equation}
\label{eq:n_electron}
\sum_{l=1}^{N_\text{basis}} f_l = N_\text{electron} .
\end{equation}

The pole expansion and selected inversion (PEXSI) method \cite{pexsi_lin_2009,pexsi_lin_2013,pexsi_lin_2014,selinv_lin_2011,pselinv_jacquelin_2016} provides an alternative way for solving the Kohn-Sham electronic structure without diagonalization. As a Fermi operator expansion (FOE) based method, PEXSI expands the density matrix in Eq. \ref{eq:density_matrix} using a $P$-term pole expansion:

\begin{equation}
\label{eq:pole}
n \approx \sum_{l=1}^{P} \Im \left( \omega^{\rho}_l (\boldsymbol{H} - (z_l + \mu) \boldsymbol{S})^{-1} \right) .
\end{equation}

\noindent Here the complex shifts $\{z_{l}\}$ and weights $\{\omega^{\rho}_l\}$ are determined through a semi-analytic formula based on contour integration, and take only a negligible amount of time to compute. The number of terms of the pole expansion is proportional to $\log(\beta\Delta E)$, where $\beta = 1/(k_B T)$ is the inverse of the thermal energy and $\Delta E$ is the spectral radius. The logarithmic scaling makes the pole expansion a highly efficient approach to expand the Fermi operator. Typically $40\sim 80$ poles are sufficient for the result obtained from PEXSI to be fully comparable ($\mu$eV/atom \cite{pexsi_lin_2013,pexsi_lin_2014}) to that obtained from diagonalization.

At first it may seem that the entire Green's function-like object $(\boldsymbol{H} - (z_l + \mu) \boldsymbol{S})^{-1}$ needs to be computed. However, if targeting at the electron density $n(\boldsymbol{r})$, in general only the entries corresponding to the non-zero pattern of $\boldsymbol{H}$ and $\boldsymbol{S}$ are actually needed. Then a selected inversion algorithm can be used to efficiently compute these selected elements of the Green's function object, and therefore the electron density.

The computational cost of the PEXSI technique scales at most as $O(N^2$). The actual complexity depends on the dimensionality of the system: $O(N)$ i.e. linear scaling for quasi-1D systems such as nanotubes; $O(N^{1.5})$ for quasi-2D systems such as surfaces and slabs; and $O(N^2)$ for general 3D bulk systems. This favorable scaling hinges on the sparse character of the Hamiltonian and overlap matrices, but not on any fundamental assumption about the localization properties of the single particle density matrix. This method is not only applicable to the efficient computation of the electron density, but also to other physical quantities such as the free energy, atomic forces, density of states and local density of states, all obtainable without computing any eigenvalues or eigenvectors \cite{pexsi_lin_2013}. These quantities can be given by pole expansions with the same complex shifts as those used for computing the electron density, with different weights.

PEXSI allows the usage of a hybrid scheme of density of states estimation based on Sylvester's law of inertia \cite{inertia_sylvester_1852}, and Newton's method to obtain the chemical potential \cite{pexsi_lin_2014}, hereafter referred to as the PEXSI mu iteration. This is an efficient and relatively robust approach with respect to the initial guess of the chemical potential, with or without the presence of gap states. A reasonable initial guess, e.g. obtained from the previous SCF step, can often converge the PEXSI mu iteration in one step.

The PEXSI method has a two-level parallelism structure and is by design highly scalable. The recently developed massively parallel PEXSI technique can make efficient use of $10,000 \sim 100,000$ processors on high performance machines.

\section{The ELSI Infrastructure}
\label{sec:elsi}
\subsection{Overview of the ELSI Interface}
\label{subsec:api_overview}
KS-DFT is implemented by a broad, diverse ecosystem of different software packages with different specialties and different numerical discretization strategies (see, e.g., Ref. \cite{dft_ghiringhelli_2016} for a listing of 46 packages). The Kohn-Sham eigenvalue problem is unavoidable in all these packages. Since the most efficient way to solve the problem may depend on factors such as system size and character (insulating or metallic), sparsity of matrices involved, density-functional employed, etc., from a user's perspective, a library that can dynamically switch between different methods according to the features of the problem is preferred. As a first step to achieve this goal (the objective of this paper), a flexible interface to different methods should enable user codes to actively select the most effective method while imposing only a minimum of format conversions, parameter tweaking, etc. on the user code.

\begin{figure*}[h!]
\centering
\includegraphics[width=0.55\textwidth]{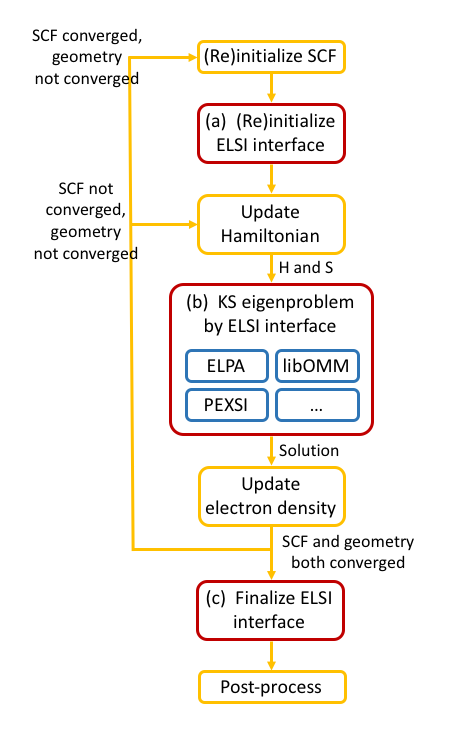}
\caption{Flow chart describing the key steps in a self-consistent field calculation based on Kohn-Sham Density-Functional Theory. Yellow boxes: Key steps commonly implemented in KS-DFT codes to perform a single SCF cycle or multiple successive SCF cycles with different atomic structures, e.g. for molecular dynamics or for geometry optimizations. Red boxes: Required additions to use the ELSI interface, including (a) initialization of the ELSI interface, (b) computing the eigensolution or the density matrix using the ELSI solvers, and (c) finalization of the ELSI interface.}
\label{fig:elsi_feature}
\end{figure*}

Although each solver library supported in ELSI maintains a limited number of well-explained Application Programming Interfaces (APIs), integrating all of them into a KS-DFT code is still a complicated, time-consuming, and error-prone task. ELSI ships a small set of APIs that are designed for rapid integration of a variety of KS electronic structure solvers into KS-DFT codes, and at the same time provides the user with hierarchical control over the interface and the solvers. There are three key steps to use ELSI, denoted by the red boxes (a), (b) and (c) in Fig. \ref{fig:elsi_feature}: (a) The ELSI interface needs to be initialized at the beginning of an SCF calculation, and potentially re-initialized if performing successive SCF cycles, e.g. for different system geometries during a molecular dynamics simulation or during a geometry optimization calculation. (b) Within the SCF cycle, ELSI serves as a bridge between the KS-DFT codes and the KS solver libraries, by taking the Hamiltonian matrix (and the overlap matrix if it exists) as input, translating the eigenproblem into a solver-specific format, invoking the solver to compute the eigenvalues and eigenvectors, or the density matrix, and finally translating the results back to the native format of the KS-DFT codes. (c) When ELSI is no longer needed, it should be finalized to deallocate any arrays internally allocated by ELSI.

\subsection{Matrix Storage and Distribution in ELSI}
\label{subsec:matrix}
The first emerging practical consideration when developing a unified software interface is the choice of matrix storage and distribution strategy. The sparsity of matrices in KS-DFT varies dramatically from small to large systems, and from 1D to 3D systems. In general, when using localized basis functions, the sparsity of matrices increases as the simulated system becomes larger. Lower dimensional systems often generate more sparse matrices. Since the effective information is only represented by the non-zero matrix elements, storing and operating on all the matrix elements lead to unnecessary memory consumption and computational complexity for very sparse matrices.

\begin{figure*}[h!]
\centering
\includegraphics[width=0.75\textwidth]{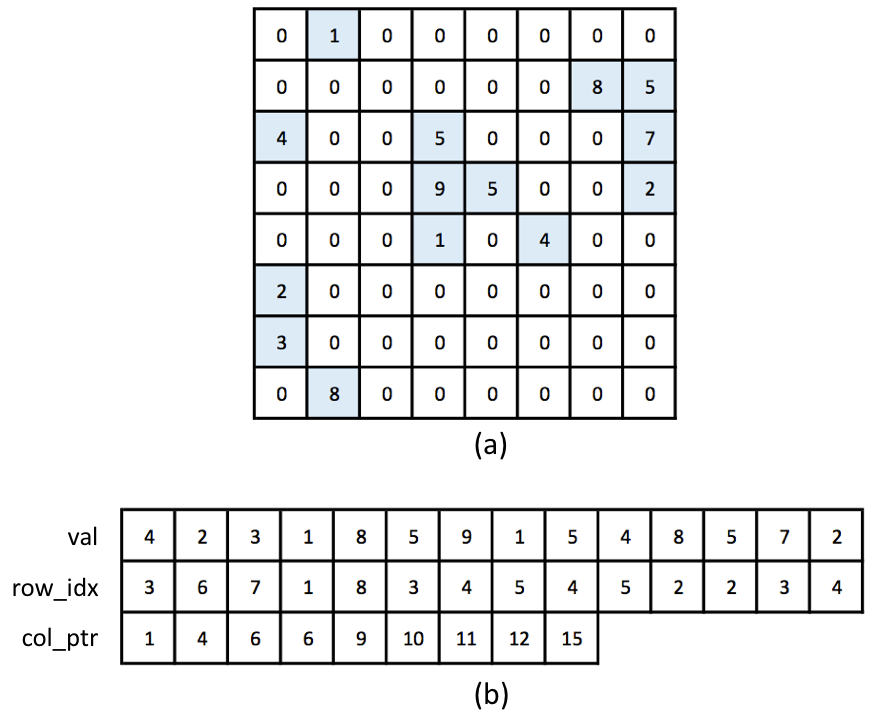}
\caption{An $8 \times 8$ matrix stored in (a) dense storage format versus in (b) compressed sparse column (CSC) storage format. In the CSC format, only the values of the non-zero elements, indicated in blue in (a), are stored in the ``val'' array. The row indexes of the non-zero elements are stored in the ``row\_inx'' array. The ``col\_ptr'' array stores the starting points of the matrix columns.}
\label{fig:dns_csc}
\end{figure*}

Implementing dense linear algebra operations, ELPA and libOMM handle matrices stored densely and distributed in a 2D block-cyclic distribution, whereas PEXSI performs sparse linear algebra with matrices stored in compressed sparse column (CSC, also known as compressed column storage, CCS) format in a 1D block distribution. These two combinations, hereafter referred to as BLACS\_DENSE and PEXSI\_CSC formats, respectively, are chosen as the input/output matrix format of the ELSI interface to bridge the needs of the solvers and of different KS-DFT codes. The comparison between dense matrix storage and CSC sparse matrix storage is illustrated in Fig. \ref{fig:dns_csc}, using an 8 $\times$ 8 matrix as an example. The dense storage keeps all the matrix elements including zeros and non-zeros. The CSC format, in contrast, drops the zeros and packs the remaining non-zeros into a 1D array, together with the row indices of the non-zero values and the starting points of the matrix columns. For a larger matrix with a higher sparsity, the CSC format will eventually consume less memory compared to the dense format.

\begin{figure*}[h!]
\centering
\includegraphics[width=0.8\textwidth]{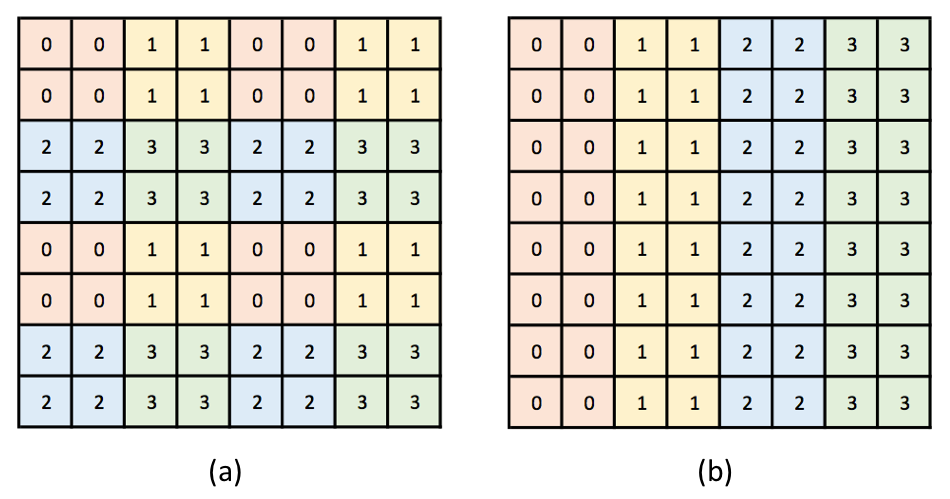}
\caption{Schematic visualizations of (a) two-dimensional block-cyclic distribution used in the BLACS\_DENSE format, and (b) one-dimensional block distribution used in the PEXSI\_CSC format, of an $8 \times 8$ matrix on 4 processors. Each unit square represents one matrix element. The integer inside each unit square denotes the index of processor where the element is stored and handled. Different processors are indicated by colors. Shown in the figure are mathematical matrices, not arrays in computers. The actual matrix storage on each processor is arbitrary, e.g. dense storage used by the BLACS\_DENSE format, CSC sparse storage used by the PEXSI\_CSC format.}
\label{fig:distributions}
\end{figure*}

To compare the two supported distributions of matrices across multiple processors in parallel computations, Fig. \ref{fig:distributions} shows how the 2D block-cyclic and the 1D block distributions are applied to the same 8 $\times$ 8 matrix. We note that shown in Fig. \ref{fig:distributions} are two mathematical matrices, the shapes of which do not represent the actual arrays in the computer. The 2D block-cyclic distribution in Fig. \ref{fig:distributions} (a) divides the global matrix into several blocks, then maps the blocks to the processors in a round-robin fashion in both the row and the column directions. The 1D block distribution in Fig. \ref{fig:distributions} (b) groups continuous matrix columns together, then linearly maps the groups of columns to the processors. In ELSI, when the input matrices are in a different distribution from the internally used one, a redistribution of the non-zero matrix elements is performed internally, i.e. no unnecessary communication of the zero elements. This redistribution is managed by the all-to-all communication implemented in the Message Passing Interface (MPI) library. Once the matrix is correctly distributed, conversion to various formats is then handled concurrently on all the MPI tasks, with each task converting a local matrix of the size at most $N^2_\text{basis}/N_\text{MPI}$, where $N_\text{MPI}$ is the number of MPI tasks involved.

\subsection{Parallelization Strategy and Interaction of ELSI with an Existing KS-DFT Code}
\label{subsec:parallel}
An important distinction in KS-DFT calculations is whether the system considered is isolated or is periodically repeated in space. In periodic systems, the full problem can be separated into subproblems defined at selected k-points in the Brillouin zone, or in a convenient unit cell in reciprocal space. The Hamilton and overlap matrices for multiple k-points are block-diagonal, such that each block on the diagonal corresponds to an eigenproblem of one k-point. These eigenproblems can therefore be solved in a embarrassingly parallel fashion side by side. For periodic systems with a small unit cell, thousands of k-points or even more can be necessary for an accurate description of the electronic structure. For a large system, in contrast, the Brillouin zone may already be well-represented by the origin of the reciprocal space known as the $\Gamma$ point.

\begin{figure*}[h!]
\centering
\includegraphics[width=0.95\textwidth]{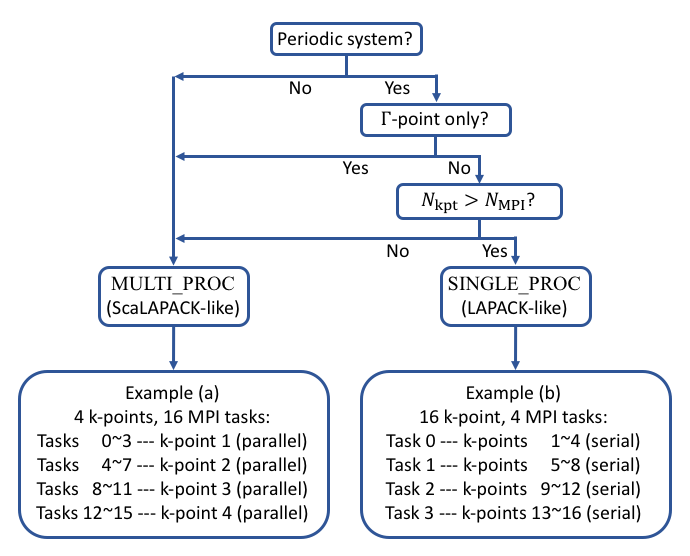}
\caption{Diagrammatic explanations of the two parallelization strategies supported by ELSI. $N_\text{kpt}$ is the number of k-points. $N_\text{MPI}$ is the number of MPI tasks.}
\label{fig:parallel}
\end{figure*}

Depending on the number of k-points $N_\text{kpt}$ (here defined to be 1 also for isolated, non-periodic cases) and the number of MPI tasks $N_\text{MPI}$, two different categories of possible KS-DFT calculations arise, as explained in Fig. \ref{fig:parallel}. Correspondingly, ELSI supports two parallelization strategies that can be specified by a parallel\_mode parameter (see also elsi\_init subroutine in Section \ref{subsec:init}):

\begin{itemize}
\item MULTI\_PROC mode, to be used if $N_\text{MPI} \geq N_\text{kpt}$. For instance, there are 4 k-points in example (a) in Fig. \ref{fig:parallel}, handled by 16 MPI tasks. The MULTI\_PROC parallelization divides the 16 MPI tasks into 4 groups. Each k-point is handled by 4 MPI tasks in the same group, and the eigenproblems of the 4 k-points are solved simultaneously by the 16 MPI tasks. Since each k-point is solved by multiple MPI tasks (processes), this parallelization mode is called MULTI\_PROC. This mode should be chosen for isolated systems, periodic systems with only one k-point, e.g. the $\Gamma$ point, and periodic systems with $N_\text{kpt}$ k-points treated by $N_\text{MPI} \geq N_\text{kpt}$ MPI tasks.

\item SINGLE\_PROC mode, to be used if $N_\text{MPI} < N_\text{kpt}$. For instance, there are 16 k-points in example (b) in Fig. \ref{fig:parallel}, handled by 4 MPI tasks. The SINGLE\_PROC parallelization divides the 16 k-points into 4 groups. Each MPI task handles 4 k-points in the same group, one after another. Since each k-point is solved by a single MPI task (process), this parallelization mode is called SINGLE\_PROC. This mode should be chosen for periodic systems with $N_\text{kpt}$ k-points treated by $N_\text{MPI} < N_\text{kpt}$ MPI tasks.
\end{itemize}

The ELPA eigensolver supported in ELSI is available for both parallel modes, returning eigensolutions for each k-point. In this case, the KS-DFT codes can then assemble the pieces of the solutions (eigenvalues and eigenvectors) returned by the solver and construct the electron density. The density matrix solvers in the 2017.05 release of ELSI do not yet support periodic calculations with more than one k-point. The ability to return combined density matrices obtained from ELPA, libOMM, or PEXSI is planned as a next step of the ELSI interface.

Once the matrix storage format and the parallel mode are decided, the usage of ELSI in KS-DFT codes becomes straightforward. Algorithm \ref{alg:elsi} summarizes in pseudo-code all the possible use cases of the ELSI interface as of the 2017.05 release. In Algorithm \ref{alg:elsi}, the main steps are denoted by subroutine names that will be systematically introduced in the following subsections. Furthermore, the initialization of the SCF calculation, updating the Hamiltonian and the electron density, checking the SCF convergence, postprocessing, and potentially further steps are all tasks that are not handled by ELSI but that are instead expected to be executed by the specific KS-DFT code calling ELSI.

Before showing detailed descriptions of the ELSI API in the next subsections, we here first introduce the concept of elsi\_handle, a Fortran derived data type containing all runtime parameters, e.g. the choices of solver, matrix\_storage\_format, and parallel\_mode (see elsi\_init subroutine in Section \ref{subsec:init}). It is intended to avoid global variables in ELSI and to allow concurrent instances of ELSI by passing around the handle as arguments. A handle can be initialized with the elsi\_init subroutine, then should be passed to all other ELSI subroutines. The ELSI interface, including the elsi\_handle, is fully interoperable with C and C++ programming languages. The elsi\_handle is defined in C/C++ as an ``opaque'' pointer, which can be seamlessly connected to a derived data type in Fortran by the iso\_c\_binding feature in Fortran compilers.

\begin{algorithm}[h!]
\caption{Usage of ELSI interface in KS-DFT codes. Pseudo-code in line 3-11, line 14-27, and line 31 corresponds to Fig. \ref{fig:elsi_feature} (a), (b), and (c), respectively.}
\begin{algorithmic}[1]
\Procedure{elsi}{}
	\State{initialize SCF calculation}
	\State{call elsi\_init}
	\If{(parallem\_mode = MULTI\_PROC)}
		\State{call elsi\_set\_mpi}
    	\If{(matrix\_storage\_format = BLACS\_DENSE)}
			\State{call elsi\_set\_blacs}
		\ElsIf{(matrix\_storage\_format = PEXSI\_CSC)}
			\State{call elsi\_set\_csc}
		\EndIf
	\EndIf
	\While{(SCF not converged)}
    	\State{update Hamiltonian}
		\State{call elsi\_customize}
		\If{(desired output: eigensolution)}
			\If{(matrix\_storage\_format = BLACS\_DENSE)}
	        	\State{call elsi\_ev\_\{real$\vert$complex\}}
			\ElsIf{(matrix\_storage\_format = PEXSI\_CSC)}
				\State{call elsi\_ev\_real\_sparse}
			\EndIf
		\ElsIf{(desired output: density matrix)}
			\If{(matrix\_storage\_format = BLACS\_DENSE)}
				\State{call elsi\_dm\_real}
			\ElsIf{(matrix\_storage\_format = PEXSI\_CSC)}
				\State{call elsi\_dm\_real\_sparse}
			\EndIf
		\EndIf
        \State{update electron density}
        \State{check SCF convergence}
	\EndWhile
	\State{call elsi\_finalize}
	\State{post-process}
\EndProcedure
\end{algorithmic}
\label{alg:elsi}
\end{algorithm}

\subsection{ELSI Initialization}
\label{subsec:init}
In this and the following subsections (Sections \ref{subsec:scf}, \ref{subsec:customize}, \ref{subsec:elsi_final}), we provide details of the capabilities of the ELSI interface as current in the 2017.05 release. Since these capabilities are intimately tied to the actual implementation, we here explain them grouped by individual subroutines as also shown in Algorithm \ref{alg:elsi}. In all instances, elsi\_h denotes the ELSI handle.

In the initialization phase, ELSI can be set up to reflect the physical quantities that usually do not change within an SCF calculation (i.e. fixed atomic structure), such as the number of basis functions and the number of electrons in the system. Implementations of SCF typically initialize these quantities before the SCF cycle begins, then keep reusing them within the cycle to repeatedly solve KS problems with an updated Hamiltonian matrix and a fixed overlap matrix. Similarly, the ELSI interface only needs to be (re-)initialized whenever the SCF cycle is itself (re-)initialized. The subroutines that are used to initialize ELSI include:

\begin{itemize}
\item \textbf{elsi\_init} (elsi\_h, solver, matrix\_storage\_format, parallel\_mode, n\_basis, n\_state, n\_electron)

(line 3 in Algorithm \ref{alg:elsi}) -- Initializes an ELSI handle with user's choices of the solver, the matrix format and distribution, the parallelization strategy, and system information including the number of basis functions, the number of eigenstates to compute, and the number of electrons.

\begin{itemize}
\item \textbf{elsi\_h} (type(elsi\_handle), output): An ELSI handle (see Section 
\ref{subsec:parallel} returned by elsi\_init subroutine. The same handle must be passed to other ELSI subroutines and be finalized when no longer needed. Multiple handles can be initialized if needed.

\item \textbf{solver} (integer, input): The choice of solver. Accepted options are 1 (ELPA), 2 (libOMM), and 3 (PEXSI).

\item \textbf{matrix\_storage\_format} (integer, input): Matrix storage and distribution of the Hamiltonian matrix, the overlap matrix, and the density matrix or the eigenvectors. Accepted options are 1 (BLACS\_DENSE) and 2 (PEXSI\_CSC) (see Section \ref{subsec:matrix}). The BLACS\_DENSE format is compatible with ELPA, libOMM, and PEXSI. If the chosen solver is PEXSI, the input matrices in the BLACS\_DENSE format are converted to PEXSI\_CSC internally, and the results in the PEXSI\_CSC format are back-converted to BLACS\_DENSE. The PEXSI\_CSC format is compatible with ELPA and PEXSI in the current release. Supporting the PEXSI\_CSC format with libOMM is on the list of features to be added in the near future.

\item \textbf{parallel\_mode} (integer, input): The choice of parallelization strategy. Accepted options are 1 (MULTI\_PROC) and 2 (SINGLE\_PROC) (see Section \ref{subsec:api_overview}). In the current release of ELSI, the SINGLE\_PROC mode is only compatible with ELPA, while the MULTI\_PROC mode supports all three solvers.

\item \textbf{n\_basis} (integer, input): Number of basis functions. This is equal to the global size of the Hamiltonian matrix, the overlap matrix, the density matrix, etc.

\item \textbf{n\_state} (integer, input): Number of states. For ELPA this is the number of eigenstates to be solved. For libOMM this must be the number of occupied states, without any fractional occupation numbers. PEXSI does not use this information.

\item \textbf{n\_electron} (integer, input): Number of electrons.
\end{itemize}

\item \textbf{elsi\_set\_mpi} (elsi\_h, mpi\_comm)

(line 5 in Algorithm \ref{alg:elsi}) -- Sets the MPI communicator to be used in the ELSI instance indicated by the handle.

\begin{itemize}
\item \textbf{mpi\_comm} (integer, input): An MPI communicator, containing an ordered group of MPI tasks, is required to use the functionalities implemented in the MPI library. The communicator assigned to an ELSI calculation can be the default global communicator of MPI, or a communicator created by the user (e.g. by calling the MPI subroutine MPI\_Comm\_Split), as long as it is compatible with the distribution of matrices.
\end{itemize}

\item \textbf{elsi\_set\_blacs} (elsi\_h, blacs\_ctxt, block\_size)

(line 7 in Algorithm \ref{alg:elsi}) -- Sets the BLACS context and the block size of the 2D block-cyclic distribution to be used in the ELSI instance indicated by the handle. Required before calling elsi\_ev\_real, elsi\_ev\_complex, and elsi\_dm\_real (see Section \ref{subsec:scf}).

\begin{itemize}
\item \textbf{blacs\_ctxt} (integer, input): A BLACS context encloses a group of processes and arranges them in a particular grid. Processes in the same context can safely communicate with each other, without worrying if the operations in one context interfere with operations in another context \cite{blacs_anderson_1991}. The ELSI interface requires the KS-DFT code to set up BLACS context(s), by calling BLACS subroutine BLACS\_Gridinit or BLACS\_Gridmap.

\item \textbf{block\_size} (integer, input): The block size parameter of the 2D block-cyclic distribution. The matrix operations inside ELSI interface, ELPA, and libOMM restrict the block sizes in the row and column directions to be the same.
\end{itemize}

\item \textbf{elsi\_set\_csc} (elsi\_h, nnz\_g, nnz\_l, n\_l\_cols, row\_idx, col\_ptr)

(line 9 in Algorithm \ref{alg:elsi}) -- Set the parameters of 1D block distributed CSC matrix storage (PEXSI\_CSC) to be used in the ELSI instance indicated by the handle. Required before calling elsi\_ev\_real\_sparse and elsi\_dm\_real\_sparse (see Section \ref{subsec:scf}).

\begin{itemize}
\item \textbf{nnz\_g} (integer, input): The global number of non-zero elements in the sparsity pattern.

\item \textbf{nnz\_l} (integer, input): The local number of non-zero elements in the sparsity pattern held by an MPI task.

\item \textbf{n\_l\_cols} (integer, input): The local number of matrix columns held by an MPI task.

\item \textbf{row\_idx} (integer, 1D array, input): The row index array of the CSC matrix storage format, containing the row index of each non-zero matrix element. An example is given in Fig. \ref{fig:dns_csc}.

\item \textbf{col\_ptr} (integer, 1D array, input): The column pointer array of the CSC matrix storage format, containing the starting point of each matrix column. An example is given in Fig. \ref{fig:dns_csc}.
\end{itemize}
\end{itemize}

The matrices that arise in KS-DFT can be either real or complex-valued. ELSI must account for these two possibilities as well. Since the real and complex arithmetic cases only differ in the data type of input/output matrices, they are not distinguishable at the initialization stage.

\subsection{Tasks during SCF}
\label{subsec:scf}
During the SCF cycle, the following tasks may be executed by ELSI solver subroutines to compute either the eigensolutions or the density matrix from the input Hamiltonian matrix (and overlap matrix, if it is not unity).

\begin{itemize}
\item \textbf{elsi\_ev\_real} (elsi\_h, ham, ovlp, eval, evec)

(line 17 in Algorithm \ref{alg:elsi}) -- Computes the eigenvalues and n\_state eigenvectors. Compatible solver: ELPA.

\begin{itemize}
\item \textbf{ham} (double precision real, 2D array, input \& output): The real-valued Hamiltonian matrix in the BLACS\_DENSE format set by subroutine elsi\_set\_blacs. This array is used for internal storage when solving the eigenproblem, and thus is destroyed on exit.

\item \textbf{ovlp} (double precision real, 2D array, input \& output): The real-valued overlap matrix in the BLACS\_DENSE format set by subroutine elsi\_set\_blacs.

A singularity check of the overlap matrix $\boldsymbol{S}$ is performed the first time elsi\_ev\_real is called. This is because the Cholesky factorization in Eq. \ref{eq:cholesky} requires $\boldsymbol{S}$ to be Hermitian positive-definite. While $\boldsymbol{S}$ in KS-DFT is guaranteed to be Hermitian by Eq. \ref{eq:ham_ovlp_integration}, the positive-definite condition can be numerically violated if the chosen basis set is large and (near-)singular, i.e. the lowest eigenvalues of $\boldsymbol{S}$ are too close to 0 (although still greater than 0). Using a near-singular basis set can lead to completely wrong and unpredictable numerical results, and thus should be avoided in general. In ELSI, this is done by computing all the eigenvalues of $\boldsymbol{S}$ and comparing them with a user-defined singularity tolerance $\tau$. The matrix is considered to be singular if it has one or more eigenvalues smaller than $\tau$. For a singular overlap matrix, the Cholesky decomposition is replaced by an eigendecomposition:

\begin{equation}
\label{eq:eigen_decomp}
\boldsymbol{S} = (\sqrt{\boldsymbol{\lambda}} \boldsymbol{x}) (\sqrt{\boldsymbol{\lambda}} \boldsymbol{x})^* = \boldsymbol{X} \boldsymbol{X}^* ,
\end{equation}

where the matrix $\boldsymbol{x}$ and the diagonal matrix $\boldsymbol{\lambda}$ contain the eigenvectors and eigenvalues of $\boldsymbol{S}$, and the matrix $\boldsymbol{X}$ is simply $\sqrt{\boldsymbol{\lambda}} \boldsymbol{x}$. By using eigendecomposition, the generalized eigenproblem is again transformed to the standard form in Eq. \ref{eq:after_cholesky}, with $\boldsymbol{\tilde{H}} = \boldsymbol{X}^{-1} \boldsymbol{H} (\boldsymbol{X}^*)^{-1}$ and $\boldsymbol{\tilde{c}} = \boldsymbol{X}^* \boldsymbol{c}$. In case that only the first $N_\text{nonsing}$ eigenvalues of $\boldsymbol{S}$ are greater than the threshold $\tau$, $\boldsymbol{X}$ correspondingly contains only the first $N_\text{nonsing}$ eigenvectors by dropping the $N_\text{basis} - N_\text{nonsing}$ eigenvectors associated with small eigenvalues. The eigenproblem transformation is still valid, however yields a smaller transformed $\boldsymbol{\tilde{H}}$ ($N_\text{nonsing} \times N_\text{nonsing}$). The solution of the transformed standard eigenproblem must be back-transformed accordingly.

On exit, ovlp is overwritten by either $\boldsymbol{L}$ in Eq. \ref{eq:cholesky} or $\boldsymbol{X}$ in Eq. \ref{eq:eigen_decomp}, depending on which transformation is used. If in the MULTI\_PROC mode, i.e. no MPI task handles more than one k-point, $\boldsymbol{L}$ or $\boldsymbol{X}$ can be stored in ovlp and efficiently reused throughout the SCF cycle. The Cholesky factorization or the eigendecomposition then only needs to be performed once. However, in the SINGLE\_PROC mode, since each MPI task handles a group of k-points in serial, 
memory constraints make it more difficult to reuse the matrices $\boldsymbol{L}$ or $\boldsymbol{X}$. In this case, the decision to either store $\boldsymbol{L}$ or $\boldsymbol{X}$, or to redo the decomposition in every SCF iteration, is up to the KS-DFT code that calls ELSI.

\item \textbf{eval} (double precision real, 1D array, output): The eigenvalues in ascending order.

\item \textbf{evec} (double precision real, 2D array, output): The real-valued eigenvectors in a matrix form in the BLACS\_DENSE format set by subroutine elsi\_set\_blacs.
\end{itemize}

\item \textbf{elsi\_ev\_complex} (elsi\_h, ham, ovlp, eval, evec)

(line 17 in Algorithm \ref{alg:elsi}) -- Same as elsi\_ev\_real, except that the Hamiltonian matrix, overlap matrix and eigenvectors are complex-valued.

\item \textbf{elsi\_ev\_real\_sparse} (elsi\_h, ham, ovlp, eval, evec)

(line 19 in Algorithm \ref{alg:elsi}) -- Computes the eigenvalues and n\_state eigenvectors. Compatible solver: ELPA.

\begin{itemize}
\item \textbf{ham} (double precision real, 1D array, input): The non-zero elements of the real-valued Hamiltonian matrix in the PEXSI\_CSC format set by subroutine elsi\_set\_csc. Inside ELSI, the input Hamiltonian matrix is converted to the BLACS\_DENSE format in every SCF iteration.

\item \textbf{ovlp} (double precision real, 1D array, input): The non-zero elements of the real-valued overlap matrix in the PEXSI\_CSC format set by subroutine elsi\_set\_csc. Inside ELSI, the input overlap matrix is converted to the BLACS\_DENSE format in the first SCF iteration. The singularity check of the overlap matrix is performed as in the elsi\_ev\_real case. Since the sparsity of the eigenproblem transformation matrix $\boldsymbol{L}$ or $\boldsymbol{X}$ is not guaranteed, the matrix $\boldsymbol{L}$ or $\boldsymbol{X}$ is stored internally in the BLACS\_DENSE format for further reuse throughout the SCF cycle.

\item \textbf{eval} (double precision real, 1D array, output): The eigenvalues in ascending order.

\item \textbf{evec} (double precision real, 2D array, output): The real-valued eigenvectors in a matrix form in the BLACS\_DENSE format. Note that the computed eigenvectors are returned in a dense format, for the reason that they are not in the same sparsity pattern of $\boldsymbol{H}$ and $\boldsymbol{S}$, or even not sparse at all.
\end{itemize}

\item \textbf{elsi\_dm\_real} (elsi\_h, ham, ovlp, den\_mat, energy)

(line 23 in Algorithm \ref{alg:elsi}) -- Computes the density matrix. Compatible solvers: ELPA, libOMM, PEXSI.

\begin{itemize}
\item \textbf{ham} (double precision real, 2D array, input \& output): The real-valued Hamiltonian matrix in the BLACS\_DENSE format set by subroutine elsi\_set\_blacs. This array is used for internal storage when computing the density matrix, and thus is destroyed on exit. If the chosen solver is PEXSI, the input Hamiltonian matrix is converted to the PEXSI\_CSC format in every SCF iteration.

\item \textbf{ovlp} (double precision real, 2D array, input \& output): The real-valued overlap matrix in the BLACS\_DENSE format set by subroutine elsi\_set\_blacs. If the chosen solver is PEXSI, the input overlap matrix is converted to the PEXSI\_CSC format in the first SCF iteration and reused throughout the SCF cycle. If the chosen solver is ELPA or libOMM, the singularity check of the overlap matrix is performed as in the elsi\_ev\_real case. The singularity check is not yet implemented for PEXSI.

\item \textbf{den\_mat} (double precision real, 2D array, output): The density matrix in the BLACS\_DENSE format set by subroutine elsi\_set\_blacs.

The chemical potential and occupation numbers must be known when ELPA is chosen to compute the density matrix following Eq. \ref{eq:density_matrix}. In ELSI, the chemical potential is found using a bisection algorithm that starts from an energy interval that includes the actual solution of the chemical potential. This is often guaranteed by using the lowest and highest eigenvalues of the system as the lower and upper bounds of the interval, and expanding the interval towards both ends if necessary. In each bisection step the number of electrons on both bounds and at the middle point of the interval is computed by Eq. \ref{eq:n_electron} (the summation becomes $\sum_{i=1}^{N_\text{kpt}} \sum_{j=1}^{N_\text{spin}} \sum_{k=1}^{N_\text{basis}}$ if including k-points and spin channels), to determine which subinterval the solution lies in. Then the interval can be repeatedly bisected until the computed number of electrons on either bound or at the middle point is sufficiently close to the actual number. During this process, the computation of occupation numbers $f_l$ requires a specific broadening scheme, which can be the Fermi broadening in Eq. \ref{eq:fermi_dirac}, or the Gaussian broadening \cite{gaussian_fu_1983}

\begin{equation}
\label{eq:gaussian}
f_l = 0.5 \cdot [1 - \text{erf} \left(\frac{\epsilon_l - \mu}{k_B T} \right)] ,
\end{equation}

where erf is the Gauss error function:

\begin{equation}
\label{eq:erf}
\text{erf} = \frac{2}{\sqrt{\pi}} \int_0^x e^{-t^2} dt .
\end{equation}

Although the error function is implemented as an intrinsic function in most programming languages, the error of each single evaluation can accumulate as a consequence of the summation in Eq. \ref{eq:n_electron}. This accumulation leads to an error on the order of $10^{-10}$ in term of the number of electrons, which is small but not negligible if the desired accuracy is on the same order. During the convergence of the SCF cycle, this small error can become more noticeable, since fluctuations of the norm of the density matrix (i.e. the system charge) will have a relatively large electrostatic effect, and can thus disturb the solution of the nonlinear fixed-point iteration scheme (e.g. Pulay mixing \cite{pulay_pulay_1980}) that is used to converge an SCF cycle. Therefore, it is useful and sometimes necessary to avoid charge fluctuations whenever possible, by ensuring an exact charge norm after the fact. In ELSI, when the accuracy of electron count can no longer be improved by bisection, then the remaining discrepancy (surplus of electrons in case of the upper bisection bound) is successively removed starting from the highest occupied KS states and proceeding to lower-lying states until the norm in Eq. \ref{eq:n_electron} is numerically exactly fulfilled.


\item \textbf{energy} (double precision real, output): The energy corresponding to the occupied eigenstates.
\end{itemize}

\item \textbf{elsi\_dm\_real\_sparse} (elsi\_h, ham, ovlp, den\_mat, energy)

(line 25 in Algorithm \ref{alg:elsi}) -- Computes the density matrix. Compatible solver: PEXSI.

\begin{itemize}
\item \textbf{ham} (double precision real, 1D array, input \& output): The non-zero elements of the real-valued Hamiltonian matrix in the PEXSI\_CSC format set by subroutine elsi\_set\_csc. This array is used for internal storage when computing the density matrix, and thus is destroyed on exit.

\item \textbf{ovlp} (double precision real, 1D array, input \& output): The non-zero elements of the real-valued overlap matrix in the PEXSI\_CSC format set by subroutine elsi\_set\_csc.

\item \textbf{den\_mat} (double precision real, 1D array, output): The non-zero elements of the density matrix in the PEXSI\_CSC format set by subroutine elsi\_set\_csc.

\item \textbf{energy} (double precision real, output): The energy corresponding to the occupied eigenstates.
\end{itemize}

\item \textbf{elsi\_collect\_pexsi} (elsi\_h, mu, e\_den\_mat, f\_den\_mat)

-- Collects additional results computed by PEXSI. Compatible solver: PEXSI.

\begin{itemize}
\item \textbf{mu} (double precision real, output): The chemical potential computed by PEXSI.

\item \textbf{e\_den\_mat} (double precision real, 1D array, output): The non-zero elements of the energy density matrix in the PEXSI\_CSC format set by subroutine elsi\_set\_csc.

\item \textbf{f\_den\_mat} (double precision real, 1D array, output): The non-zero elements of the free energy density matrix in the PEXSI\_CSC format set by subroutine elsi\_set\_csc.
\end{itemize}
\end{itemize}

\subsection{ELSI Customization Options}
\label{subsec:customize}
Although ELSI sets reasonable default runtime parameters for each solver whenever possible, no set of parameters can adequately cover all use cases. The elsi\_customize subroutines allow a user to determine runtime parameters explicitly, thus providing maximum flexibility to control the particulars of ELSI. Designed with the feature of optional arguments in Fortran, the elsi\_customize subroutines have a general calling syntax:

call elsi\_customize(elsi\_h, keyword=choice),

\noindent where elsi\_h is the ELSI handle to be customized, ``keyword'' is the parameter to be customized, and ``choice'' is the value to overwrite the default value of ``keyword''. Calling elsi\_customize (line 14 in Algorithm \ref{alg:elsi}) only modifies the parameter associated with elsi\_h, instead of changing the behavior of all handles.

\begin{itemize}
\item \textbf{elsi\_customize} (elsi\_h, keyword=choice)

The following customizable keywords are particularly important:

\begin{itemize}
\item \textbf{overlap\_is\_unit} (logical, input): ELSI by default assumes that the KS eigenproblem is a generalized problem (Eq. \ref{eq:generalized_evp_matrix}). Setting the keyword overlap\_is\_unit to true allows the usage of ELSI for a standard eigenproblem, e.g. when using orthonormal basis sets, or the generalized eigenproblem has been transformed to the standard form by the calling code itself. If overlap\_is\_unit is true, the singularity check for the overlap matrix described in Section \ref{subsec:scf} will be completely ignored.

\item \textbf{zero\_threshold} (double precision real, input): Threshold to define ``zero'' in ELSI matrix format conversions. When converting a dense matrix into a sparse format, any double precision number smaller than this threshold is overwritten by 0.

\item \textbf{no\_singularity\_check} (logical, input): The singularity check of the overlap matrix can be skipped here.

\item \textbf{singularity\_tolerance} (double precision real, input): The tolerance of basis singularity $\tau$ in the singularity check.
\end{itemize}

\item \textbf{elsi\_customize\_mu} (elsi\_h, keyword=choice)

Customizes the chemical potential and occupation number computation in ELSI. Customizable keywords include:

\begin{itemize}
\item \textbf{broadening\_scheme} (integer, input): The broadening scheme to be used in the determination of occupation numbers and chemical potential. Accepted options are 1 (Gaussian broadening), 2 (Fermi broadening), 3 (0$^\text{th}$ order Methfessel-Paxton broadening), and 4 (1$^\text{st}$ order Methfessel-Paxton broadening).

\item \textbf{broadening\_width} (double precision real, input): The broadening width parameter ($k_B T$ in Eq. \ref{eq:fermi_dirac} and \ref{eq:gaussian}).

\item \textbf{occ\_accuracy} (double precision real, input): Desired accuracy in terms of the sum of occupation numbers, i.e. the number of electrons, in the determination of occupation numbers and chemical potential.

\item \textbf{mu\_max\_steps} (integer, input): Maximum steps of the bisection algorithm (described as a part of subroutine elsi\_dm\_real) to compute the occupation numbers and chemical potential.
\end{itemize}

\item \textbf{elsi\_customize\_elpa} (elsi\_h, keyword=choice)

Customizes the ELPA solver. Customizable keywords include:

\begin{itemize}
\item \textbf{elpa\_solver} (integer, input). The choice of ELPA solvers. Accepted options are 1 (ELPA 1-stage solver) and 2 (ELPA 2-stage solver).
\end{itemize}

\item \textbf{elsi\_customize\_omm} (elsi\_h, keyword=choice)

Customizes the libOMM solver. Customizable keywords include:

\begin{itemize}
\item \textbf{omm\_flavor} (integer, input): The choice of method to perform OMM minimization. Accepted options are 0 (the basic flavor that follows Eq. \ref{eq:omm_energy} exactly) and 2 (the Cholesky flavor that transforms the generalized eigenproblem to the standard form using Cholesky factorization before minimization).

\item \textbf{n\_elpa\_steps} (integer, input): ELPA can be employed in the first n\_elpa\_steps SCF iterations, as these take the longest time to converge with iterative methods. Starting from the $\text{(n\_elpa\_steps + 1)}^\text{th}$ SCF step, the libOMM solver will be used with the eigenvectors computed by ELPA in the $\text{(n\_elpa\_steps)}^\text{th}$ SCF step as the initial guess for the coefficients of Wannier functions.

\item \textbf{omm\_tolerance} (double precision real, input): The stop criterion of the OMM energy functional minimization in Eq. \ref{eq:omm_energy}. This minimization is considered to be converged when the relative energy difference between subsequent line searches given by $2 (E[\boldsymbol{W}_1] - E[\boldsymbol{W}_0])/(E[\boldsymbol{W}_1] + E[\boldsymbol{W}_0])$ is smaller than or equal to this dimensionless value.
\end{itemize}

The convergence rate of the OMM energy functional minimization depends heavily on the minimization method and the initial guess of the coefficients of the Wannier functions. The effects of omm\_flavor and n\_elpa\_steps on the performance of OMM are investigated and reported in Section \ref{subsec:results_omm}.

\item \textbf{elsi\_customize\_pexsi} (elsi\_h, keyword=choice)

Customizes the PEXSI solver. Customizable keywords include:

\begin{itemize}
\item \textbf{n\_poles} (integer, input): The number of poles in the Fermi operator expansion, i.e. $P$ in Eq. \ref{eq:pole}. The pole expansion is an exact algorithm if the number of poles is infinitely large. In practice, $40 \sim 80$ poles are usually sufficient for the result obtained from PEXSI to be fully comparable to that obtained from diagonalization. Performing a convergence test with increasing number of poles is a practical approach to estimate the optimal number of poles for a KS-DFT code.

\item \textbf{n\_electron\_accuracy} (double precision real, input): The desired accuracy in term of the number of electrons out of the density matrix approximated by Eq. \ref{eq:pole}.

\item \textbf{temperature} (double precision real, input): The physical meaning of the temperature here is the energy $\beta = K_B T$ in Eq. \ref{eq:fermi_dirac}, i.e. the broadening width.

\item \textbf{delta\_e} (double precision real, input): The upper bound for the spectral radius $\Delta E$ of $S^{-1}H$. This parameter and the $\beta$ parameter affect the number of terms of the pole expansion.

\item \textbf{max\_iteration} (integer, input): The maximum number of PEXSI mu iterations to determine the chemical potential.

\item \textbf{mu\_0, mu\_min, mu\_max} (double precision real, input): The initial guess, lower bound, and upper bound for the chemical potential. A good initial guess significantly accelerates the convergence of the PEXSI mu iteration. An estimate of the chemical potential is available in PEXSI via the inertia counting procedure based on Sylvester's law of inertia. Starting from the second SCF iteration, if the change in chemical potential from the previous SCF step to the current step is small, ELSI will automatically skip the inertia counting and use the chemical potential from the previous step as the initial guess for the current step.

\item \textbf{mu\_safeguard} (double precision real, input): A fail-safe approach designed for the PEXSI mu iteration. If the error in the chemical potential computed by PEXSI is larger than this safeguard, the code will exit the mu iteration and re-invoke the inertia counting to estimate the chemical potential.
\end{itemize}
\end{itemize}

\subsection{ELSI Finalization}
\label{subsec:elsi_final}
\begin{itemize}
\item \textbf{elsi\_finalize} (elsi\_h)

(line 31 in Algorithm \ref{alg:elsi}) -- Terminates the ELSI instance associated with the handle. This deallocates any arrays internally allocated by ELSI.

\begin{itemize}
\item \textbf{elsi\_h} (type(elsi\_handle), input \& output): On exit, all the parameters of this handle are reset to ``UNSET'' or their default values. To become valid again, the handle must be re-initialized by elsi\_init.
\end{itemize}
\end{itemize}

\subsection{ELSI Software in Practice}
\label{subsec:software}
The 2017.05 release of the ELSI software package, available on the ``ELSI Interchange'' website (http://elsi-interchange.org), contains the ELSI interface described in Section \ref{sec:elsi}, as well as redistributed source code of the three solver libraries ELPA (version 2016.11.001.pre, http://elpa.mpcdf.mpg.de), libOMM (version 0.0.1, http://esl.cecam.org/LibOMM), and PEXSI (version 0.10.2, http://pexsi.org). They are redistributed with ELSI for an optional integrated installation managed by a unified make-based build system with specific keywords set by the users in ``make.sys'' files. While we focus more on the development of a unified interface to connect the KS solvers and the KS-DFT codes, the ELPA, libOMM, and PEXSI solvers themselves are being actively developed by their own communities. The three solvers linked into ELSI can be either the built-in versions shipped with ELSI, or independently built versions, e.g. pre-installed and optimized versions available on a given supercomputer. There are two external dependencies that must be downloaded and installed separately: the ParMETIS library (http://glaros.dtc.umn.edu/gkhome/metis/parmetis/overview) and the SuperLU\_DIST library (http://crd-legacy.lbl.gov/$\sim$xiaoye/SuperLU).

ELSI can be integrated directly into relevant pieces of KS-DFT codes written in Fortran, C, or C++. So far, ELSI has been tested in the DGDFT \cite{dgdft_hu_2015}, FHI-aims \cite{fhiaims_blum_2009}, NWChem \cite{nwchem_valiev_2010} (via Global Arrays Toolkit \cite{ga_nieplocha_2006}), and SIESTA \cite{siesta_soler_2002} software packages. Detailed instructions on how to obtain, install, and use the ELSI software are documented in the ELSI User's Guide \cite{elsi_yu_2017}.

\section{Benchmarks and Discussions}
\label{sec:results}
In the final part of this work, we present a comparative study of the three KS electronic structure solvers ELPA, libOMM, and PEXSI, as currently supported by ELSI. This study employs a consistent set of systems and settings, and illustrates the optimal choice of solver strategies in different scenarios and system size ranges. The Hamiltonian and overlap matrices are constructed from actual DFT-PBE \cite{pbe_perdew_1996} calculations using the all-electron, full-potential electronic structure code FHI-aims (Fortran) with a ``tier 1'' numeric atom-centered orbital (NAO) basis set \cite{fhiaims_blum_2009,nao_havu_2009}, and the pseudopotential code DGDFT (C++) with an adaptive local basis (ALB) \cite{alb_lin_2012,dgdft_hu_2015}. Both packages have been demonstrated to perform large-scale DFT calculations with at least thousands of atoms \cite{elpa_marek_2014,dgdft_hu_2015,sicgraphene_nemec_2013}. Details of the KS-DFT code specific settings are given in \ref{app:fhiaims} and \ref{app:dgdft}, respectively. As the benchmark systems, we selected 2D graphene supercell models with sizes ranging from 1,800 to 11,520 atoms. All calculations reported here are $\Gamma$-point-only (the ELSI interface is thus in MULTI\_PROC mode) and real arithmetic. Among the benchmark problems, the graphene $30 \times 30 \times 1$, $45 \times 45 \times 1$, and $60 \times 60 \times 1$ supercell models have a small band gap of about 0.002 meV, since the Dirac cone of graphene, whose coordinates in the reciprocal space are (1/3, 1/3, 0), is included in the folded images of the $\Gamma$ point. The other graphene models have a band gap of 0.34 $\sim$ 0.51 eV. The dimensions of the models, the number of employed basis functions, and the sparsity factor of the corresponding matrices are reported in Table \ref{tab:test_system}. The maximum differences in the converged total energies are 6.3 $\mu$eV/atom between the results obtained with ELPA and libOMM, and 0.8 $\mu$eV/atom between ELPA and PEXSI. We note that separate benchmarks of ELPA, libOMM, and PEXSI applied to insulating/semiconducting, 1D/3D systems have been reported in earlier publications \cite{elpa_auckenthaler_2011,elpa_marek_2014,libomm_corsetti_2014,pexsi_lin_2009,pexsi_lin_2013}.

We here report, to our knowledge, the first directly comparable benchmark of all three approaches for the same system and using exactly the same hardware and software environment. All computations were performed on the Cray XC30 supercomputer Edison at National Energy Research Scientific Computing Center (NERSC). Each node of Edison is equipped with two 12-core Intel Ivy Bridge processors. The nodes were fully exploited by launching 24 MPI tasks on each node. No multi-threading parallelization was employed.

\begin{table*}[h!]
\centering
\caption{Supercell size, number of atoms $N_\text{atom}$, number of basis functions $N_\text{basis}$, and sparsity factor $N_\text{zero}/N^2_\text{basis}$ of the graphene systems used in this work. $N_\text{zero}$ is the number of zero elements in the Hamiltonian matrices. FHI-aims models contain 2 carbon atoms in each unit cell, and results are shown in Figs. \ref{fig:sovlers}, \ref{fig:redist}, \ref{fig:init}, \ref{fig:elpa_steps}, \ref{fig:omm_flavors}, \ref{fig:pexsi_steps}, and \ref{fig:fhiaims}. DGDFT models contain 10 graphene layers (20 carbon atoms) in each unit cell, and results are shown in Fig. \ref{fig:pexsi_scaling}.}
\begin{tabular}{c c c c c c c}
\hline
\hline
Code & Model & Supercell & $N_\text{atom}$ & $N_\text{basis}$ & $N_\text{zero}/N^2_\text{basis}$ \\
\hline
FHI-aims & Graphene & $30 \times 30 \times 1$ &  1800 &  25200 & 97.50\% \\
FHI-aims & Graphene & $35 \times 35 \times 1$ &  2450 &  34300 & 98.16\% \\
FHI-aims & Graphene & $40 \times 40 \times 1$ &  3200 &  44800 & 98.58\% \\
FHI-aims & Graphene & $45 \times 45 \times 1$ &  4050 &  56700 & 98.88\% \\
FHI-aims & Graphene & $50 \times 50 \times 1$ &  5000 &  70000 & 99.09\% \\
FHI-aims & Graphene & $55 \times 55 \times 1$ &  6050 &  84700 & 99.25\% \\
FHI-aims & Graphene & $60 \times 60 \times 1$ &  7200 & 100800 & 99.41\% \\
\hline
DGDFT    & Graphene & $18 \times 18 \times 1$ &  6480 &  97200 & 99.98\% \\
DGDFT    & Graphene & $24 \times 24 \times 1$ & 11520 & 172800 & 99.99\% \\
\hline
\hline
\end{tabular}
\label{tab:test_system}
\end{table*}

\subsection{Performance of the ELPA, libOMM, PEXSI Solvers}
\label{subsec:results_solvers}
We first compare the performance of the key computational steps of the ELSI solvers that are repeated in every SCF iteration. These repeated steps are: transforming the eigenproblem (Eq. \ref{eq:after_cholesky}), solving the standard eigenproblem (Fig. \ref{fig:elpa2}), and back-transforming the eigenvectors in ELPA; the minimization (CG line search) of OMM energy functional (Eqs. \ref{eq:reduced_ham_ovlp} and \ref{eq:omm_energy}), and the construction of density matrix from the final Wannier functions in libOMM; the numerical factorization and the selected inversion of the object $\boldsymbol{H} - (z_l + \mu) \boldsymbol{S}$ (Eq. \ref{eq:pole}), and the construction of density matrix from the poles in PEXSI. There are other computationally expensive steps that only occur in the first SCF iteration and have less significant effects on the total time of an SCF cycle. The performance of those steps is discussed separately in Sections \ref{subsec:results_format} and \ref{subsec:results_init}. For reference, the performance of the remaining computational steps (in addition to the KS eigenproblem) of standard DFT-PBE calculations using FHI-aims code is shown in \ref{app:fhiaims}.

\begin{figure*}[h!]
\centering
\includegraphics[width=\textwidth]{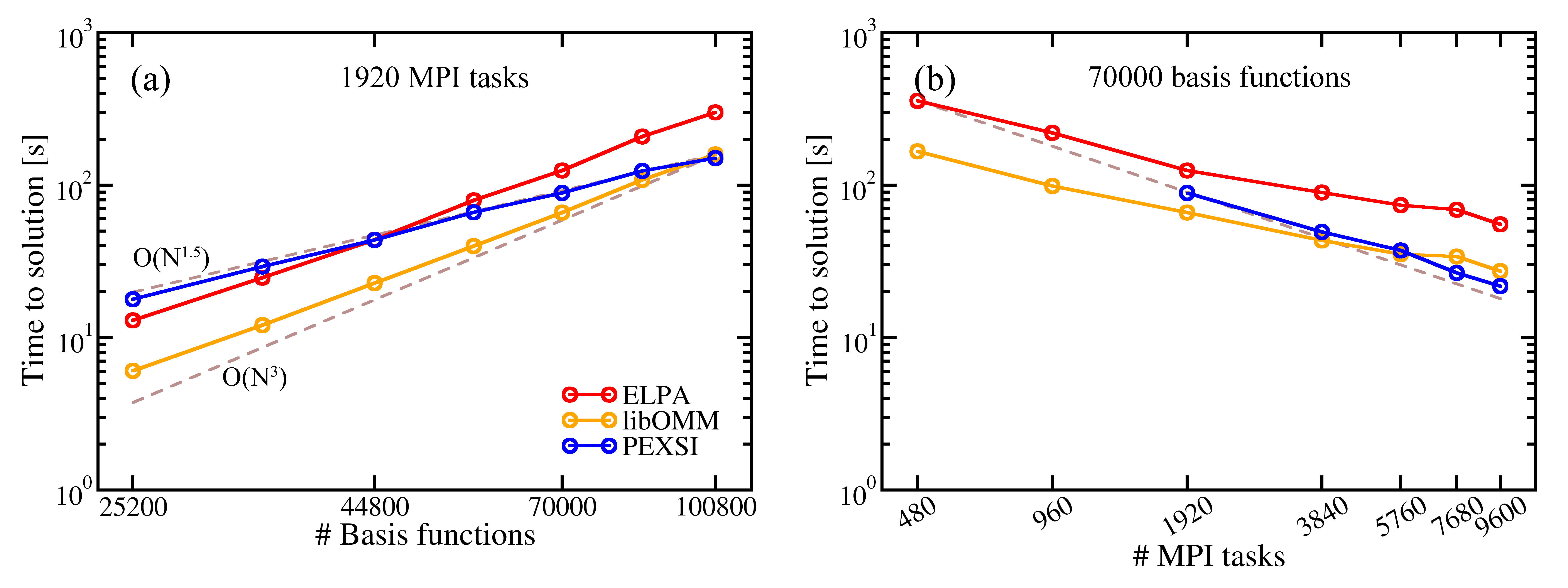}
\caption{Scaling of the ``repeated'' steps in ELPA, libOMM, and PEXSI solvers with respect to (a) the number of basis functions and (b) the number of MPI tasks. The number of MPI tasks in (a) is 1,920. The number of basis functions in (b) is 70,000. The ``repeated'' steps are: transforming the eigenproblem (Eq. \ref{eq:after_cholesky}), solving the standard eigenproblem (Fig. \ref{fig:elpa2}), and back-transforming the eigenvectors in ELPA; the minimization of OMM energy functional (Eqs. \ref{eq:reduced_ham_ovlp} and \ref{eq:omm_energy}), and the construction of density matrix from the final Wannier functions in libOMM (the CG line search converges in one step); the numerical factorization and the selected inversion of the object $\boldsymbol{H} - (z_l + \mu) \boldsymbol{S}$ (Eq. \ref{eq:pole}), and the construction of density matrix from the poles in PEXSI (the PEXSI iteration converges in one step). Ideal scaling is indicated by the dashed lines. PEXSI cannot solve the problem of 70,000 basis functions (5,000 carbon atoms) with 480 or 960 MPI tasks, due to the limited amount of memory assigned to each pole.}
\label{fig:sovlers}
\end{figure*}

Fig. \ref{fig:sovlers} shows the wall clock time of the above-mentioned repeated steps of the solvers. It is worth noting that, when using the same computational resources, the time used by ELPA is theoretically constant during an SCF cycle, as the performance of a dense direct eigensolver only depends on the size of the matrix to solve. In contrast, the time used by libOMM and PEXSI depends on the number of CG line searches and the number of PEXSI mu iterations, respectively. Since both the number of CG line searches and the number of PEXSI mu iterations can be quickly reduced to 1 as the SCF cycle proceeds, shown in Fig. \ref{fig:sovlers} are the timings corresponding to 1 CG line search in libOMM using the basic flavor (see Section \ref{subsec:results_omm} for the effect of flavor on the performance of libOMM), and 1 PEXSI mu iteration in PEXSI. In future versions of PEXSI, a newly designed algorithm will be used to update the chemical potential as the SCF cycle converges, and the number of mu iterations will always be 1 in each SCF iteration.

In Fig. \ref{fig:sovlers} (a), the scaling of solvers with respect to the basis size is shown for DFT-PBE calculations of graphene models consisting of 1,800 atoms (25,200 basis functions) to 7,200 atoms (100,800 basis functions) using 1,920 MPI tasks. Both ELPA and libOMM exhibit scalings close to $O(N^3)$, as expected. In this particular set-up, libOMM is consistently faster than ELPA by a factor of 2. PEXSI, with a lower computational complexity (theoretically $O(N^{1.5})$ for 2D systems), begins to outperform ELPA and libOMM at around 3,000 atoms and 7,000 atoms, respectively. The benefit of using PEXSI should become more significant as we further increase the system size.

The strong scaling shown in Fig. \ref{fig:sovlers} (b) demonstrates the scalability of the solvers when they are applied to the graphene 5,000-atom model (70,000 basis functions) using 480 to 9,600 MPI tasks. All three solvers exhibit good scalability to 9,600 MPI tasks. In particular, the PEXSI solver scales almost ideally up to thousands of MPI tasks. This is attributed to the 2-level parallelism employed in PEXSI (Section \ref{subsec:pexsi_intro}). The perfect strong scaling of PEXSI can be further extended to at least tens of thousands of MPI tasks (this is demonstrated in Section \ref{subsec:results_pexsi}). However, PEXSI fails to solve the problem with 480 or 960 MPI tasks, owing to the limited memory assigned to each pole.

\subsection{Matrix Redistribution}
\label{subsec:results_format}
\begin{figure*}[h!]
\centering
\includegraphics[width=\textwidth]{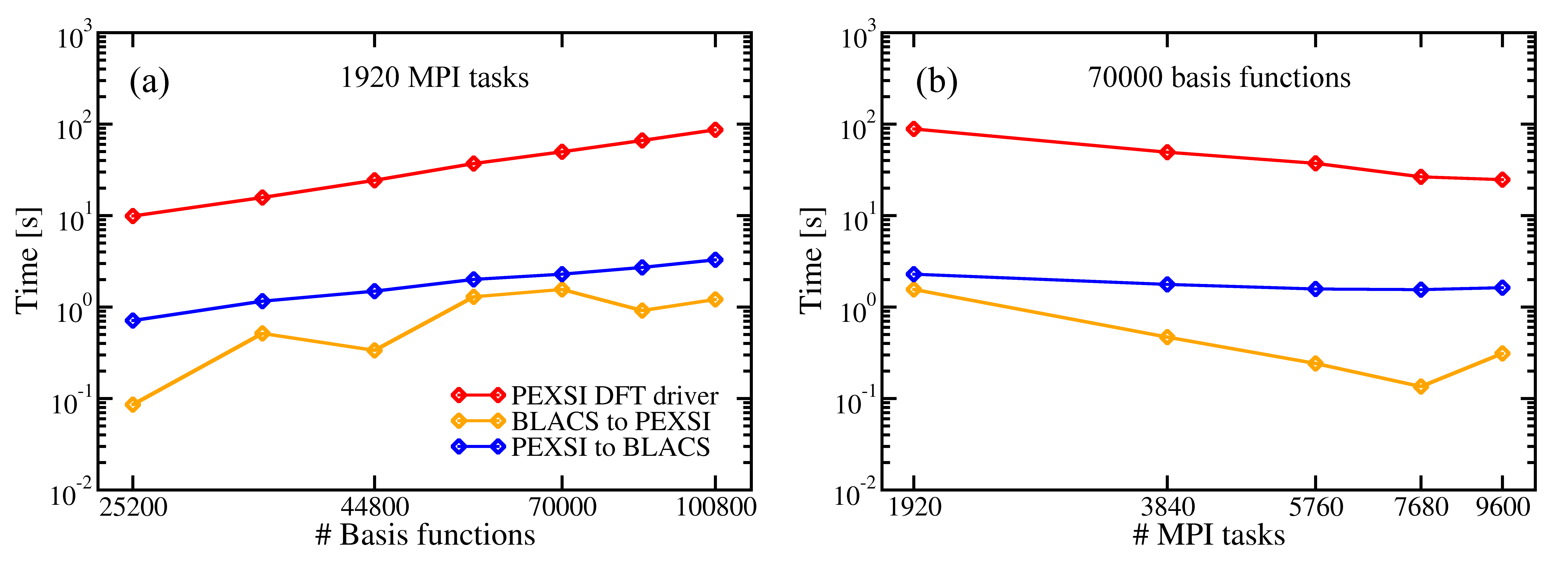}
\caption{Scaling of matrix redistribution with respect to (a) the number of basis functions and (b) the number of MPI tasks. The number of MPI tasks in (a) is 1,920. The number of basis functions in (b) is 70,000. BLACS to PEXSI: redistribution of the Hamiltonian matrix from 2D block-cyclic dense storage (BLACS\_DENSE) to 1D block CSC sparse storage (PEXSI\_CSC). PEXSI to BLACS: redistribution of the density matrix from 1D block CSC sparse storage (PEXSI\_CSC) to 2D block-cyclic dense storage (BLACS\_DENSE). The overlap matrix is redistributed only once per SCF cycle, hence its absence here.}
\label{fig:redist}
\end{figure*}

When using the elsi\_dm\_real subroutine (Section \ref{subsec:scf}) to compute the density matrix with the BLACS\_DENSE format and the PEXSI solver, the input Hamiltonian and overlap matrices are not in the correct format for PEXSI. The elsi\_dm\_real subroutine internally converts the input Hamiltonian matrix to the PEXSI\_CSC format, and converts the density matrix computed by PEXSI back to the original format. The overlap matrix is converted as well, albeit only in the first iteration of an SCF cycle. The performance of the Hamiltonian matrix conversion from BLACS\_DENSE to PEXSI\_CSC and the density matrix conversion from PEXSI\_CSC to BLACS\_DENSE are shown and compared to the PEXSI computation time in Fig. \ref{fig:redist}. For matrix sizes ranging from 25,200 (1,800 atoms) to 100,800 (7,200 atoms), Fig. \ref{fig:redist} (a) shows that the wall clock time for both conversions with 1,920 MPI tasks is always below 10\% of the PEXSI computation time (red lines in Fig. \ref{fig:redist}). Fig. \ref{fig:redist} (b) shows that the data redistribution time is consistently below 10\% of the computation time, when using 1,920 to 9,600 MPI tasks for a problem of dimension 70,000. The BLACS\_DENSE to PEXSI\_CSC conversion stops scaling at 9,600 MPI tasks. Further optimization of the conversion using MPI point-to-point communications is planned as a future work direction.

\subsection{SCF Initialization}
\label{subsec:results_init}
Computational steps that are only required in the first one or few SCF iterations have some impact on the overall performance of an SCF cycle. Here we discuss three such steps: (1) Cholesky factorization of the overlap matrix in Eq. \ref{eq:cholesky}, which is used to transform the generalized eigenvalue problem to the standard form. This is a mandatory step for ELPA and an optional step for libOMM. The Cholesky factorization of a dense matrix in ELSI is performed using subroutines provided in ELPA. (2) Symbolic factorization that provides PEXSI necessary information of the sparsity pattern of the Hamiltonian and overlap matrices before numerical factorization and selected inversion are carried out. The symbolic factorization of a sparse matrix is performed using subroutines provided in the SuperLU\_DIST library \cite{superlu_li_2003,superlu_grigori_2007}. (3) Inertia counting that quickly estimates the chemical potential of the system according to Sylvester's Inertia Law theorem \cite{inertia_sylvester_1852}. This reasonable initial guess of the chemical potential is essential to the fast convergence of PEXSI.

\begin{figure*}[h!]
\centering
\includegraphics[width=\textwidth]{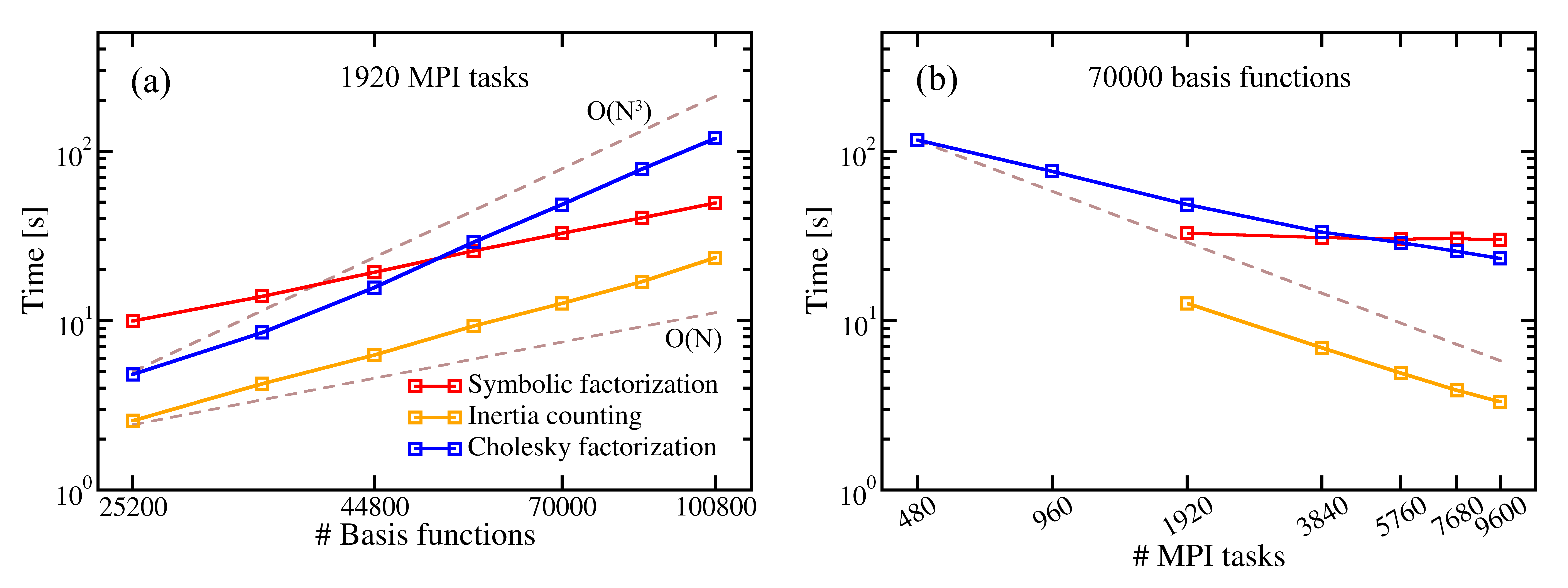}
\caption{Scaling of symbolic factorization using SuperLU\_DIST, inertia counting using PEXSI, and Cholesky factorization using ELPA, with respect to (a) the number of basis functions and (b) the number of MPI tasks. The number of MPI tasks in (a) is 1,920. The number of basis functions in (b) is 70,000. Symbolic factorization is performed in serial. Ideal scaling is indicated by the dashed lines.}
\label{fig:init}
\end{figure*}

Fig. \ref{fig:init} (a) shows the wall clock time of the three initialization steps as a function of the system size. The Cholesky factorization of a dense matrix using ELPA subroutines scales cubically with the system size, whereas the symbolic factorization and inertia counting scale linearly. The scaling difference among these preprocessing steps helps explain why PEXSI is more favorable for large systems. In the strong scaling plot shown in Fig. \ref{fig:init} (b), the dense Cholesky factorization is shown to scale up to 9,600 MPI tasks. Because the symbolic factorization implemented in SuperLU\_DIST is not stable when executed on multiple processors, we used a sequential version of the symbolic factorization in the experiment, which obviously does not scale. We are in the process of developing a more robust and scalable implementation of the symbolic factorization procedure as part of the development of a new parallel sparse Cholesky (and LDLT) factorization library called symPACK \cite{sympack_jacquelin_2016}.

\subsection{ELPA}
\label{subsec:results_elpa}
\begin{figure*}[h!]
\centering
\includegraphics[width=\textwidth]{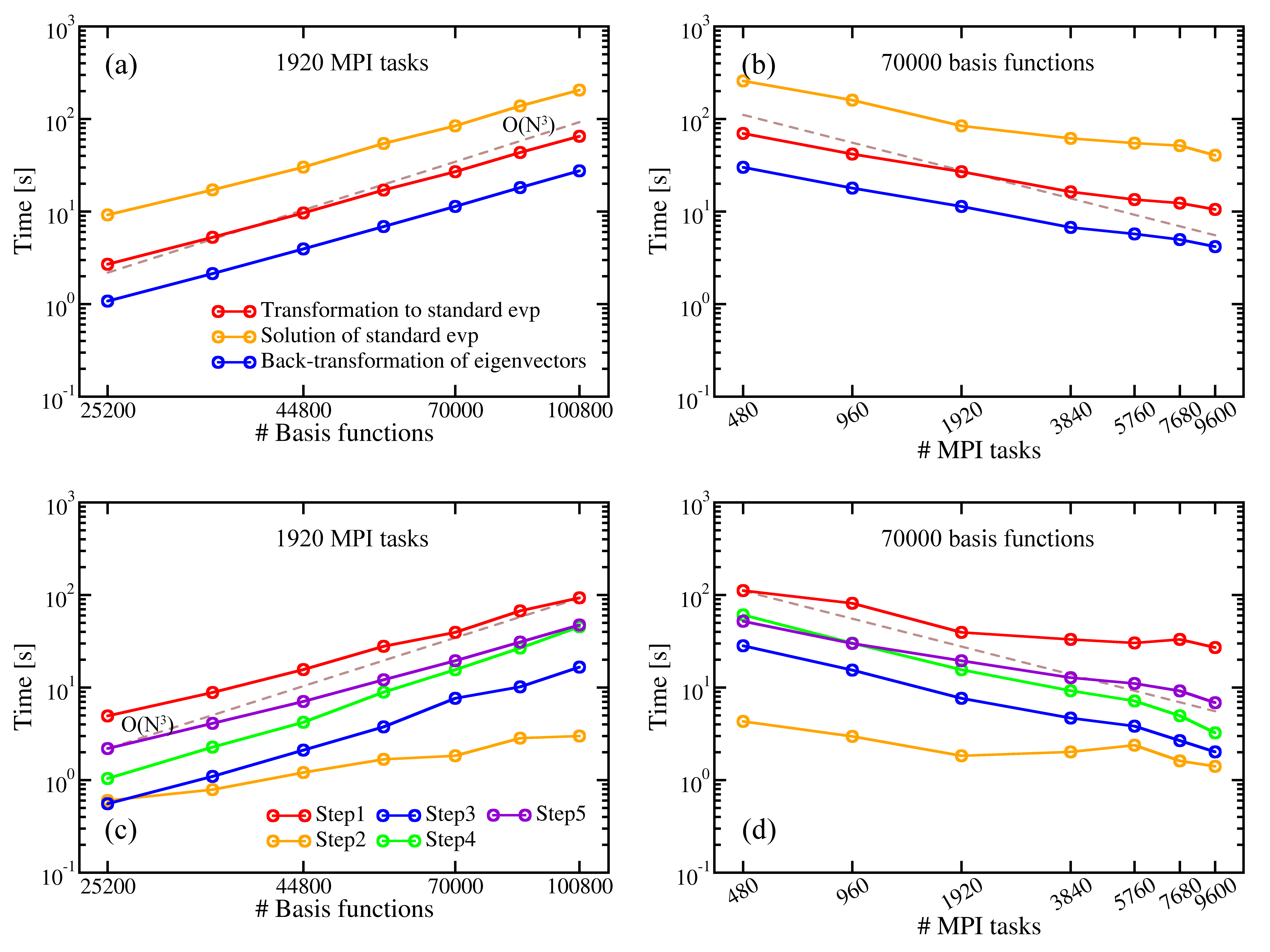}
\caption{Scaling of the key computational steps of the ELPA eigensolver with respect to (a,c) the number of basis functions and (b,d) the number of MPI tasks. The number of MPI tasks in (a) is 1,920. The number of basis functions in (b) is 70,000. Ideal scaling is indicated by the dashed lines. The upper panel (a,b) focuses on the transformation from a generalized eigenproblem to its standard form, the solution of a standard problem, and the back-transformation of the eigenvectors to the original generalized problem. The lower panel (c,d) further decomposes the solution of a standard eigenproblem using the ELPA 2-stage solver into 5 substeps, as illustrated in Fig. \ref{fig:elpa2}.}
\label{fig:elpa_steps}
\end{figure*}

To analyze the performance of the ELPA eigensolver for the graphene problem solved here, the solution of a generalized eigenproblem (red lines in Fig. \ref{fig:sovlers}) is divided into three steps: the transformation of the generalized eigenproblem to the standard form (Eq. \ref{eq:after_cholesky}), the solution of the standard eigenproblem (Fig. \ref{fig:elpa2}), and the back-transformation of the eigenvectors. Fig. \ref{fig:elpa_steps} (a) and (b) show the scaling of the three steps with respect to the number of basis functions and the number of MPI tasks, respectively. All these steps scale cubically with respect to the system size. Solving the standard eigenproblem is more expensive than the transformation steps. While the three steps show similar strong scaling up to 9,600 MPI tasks, the solution of a standard eigenproblem dominates the total computation time. In Fig. \ref{fig:elpa_steps} (c) and (d), the solution time of a standard eigenproblem using ELPA 2-stage solver is further decomposed into five steps illustrated in Fig. \ref{fig:elpa2}. These plots show that the current bottlenecks in terms of both computation time and parallel efficiency are the first step, i.e. the transformation of a full matrix to a banded form, and the fifth step, i.e. the back-transformation of the eigenvectors from a banded form to a full form. The fourth step, back-transformation of the eigenvectors to the banded form, is not the most time consuming step of the computation. In fact, the computational complexity of the third, fourth, and fifth steps is roughly proportional to the number of eigenvectors to compute, as only these eigenvectors need to be calculated in the third step and transformed in the fourth and fifth steps.

\subsection{libOMM}
\label{subsec:results_omm}
The performance of the iterative OMM method depends significantly on the convergence rate of the CG minimization. The prototype OMM implementation in libOMM generates random numbers as the initial guess for the coefficients of Wannier functions used in the first SCF iteration, consequently leading to a large and unpredictable number of iterations in the CG line search scheme. Then, the convergence of line search is dramatically accelerated as the SCF cycle proceeds, as the Wannier functions coefficients calculated in the current iteration are reused as the initial guess in the next iteration. Inspired by the connection between the Wannier functions and the basis functions in Eq. \ref{eq:wannier}, a better idea is to use the eigenfunctions corresponding to the occupied space computed by ELPA as the initial guess for OMM. In ELSI, this is achieved automatically, controlled by the n\_elpa\_steps parameter (see Section \ref{subsec:customize}). Table \ref{tab:omm_cg} reflects how n\_elpa\_steps affects the CG convergence of OMM in the $\text{(n\_elpa\_steps + 1)}^\text{th}$ SCF iteration, by showing the number of CG line searches in the basic and Cholesky flavors of OMM as a function of the number of ELPA steps. In general, more ELPA steps lead to faster CG convergence. In this particular test case with 5,000 carbon atoms and 70,000 basis functions, 6 ELPA steps are sufficient to reduce the number of CG line searches in libOMM to 1 for both tested flavors.

\begin{table*}[h!]
\centering
\caption{Number of conjugate gradient (CG) line search steps required by libOMM to minimize the OMM energy functional. The benchmark system here is the graphene $50 \times 50 \times  1$ supercell model containing 5,000 atoms and 70,000 basis functions. In the table, ``Basic'' refers to the method that directly operates on the generalized eigenproblem; ``Cholesky'' refers to the method that applies Cholesky factorization to transform the generalized problem to the standard form. ``x'' in the second column means that the minimization cannot converge within the maximum allowed number of CG iterations (5000).}
\begin{tabular}{l c c c c c c c c}
\hline
\hline
\# ELPA steps    &   0 &   1 &   2 &   3 &   4 &   5 &   6 &   7 \\
\hline
\# CG (Basic)    &   x & 185 & 255 & 153 &  72 &   7 &   1 &   1 \\
\# CG (Cholesky) & 254 &  27 &  36 &  24 &  13 &   5 &   1 &   1 \\
\hline
\hline
\end{tabular}
\label{tab:omm_cg}
\end{table*}

\begin{figure*}[h!]
\centering
\includegraphics[width=\textwidth]{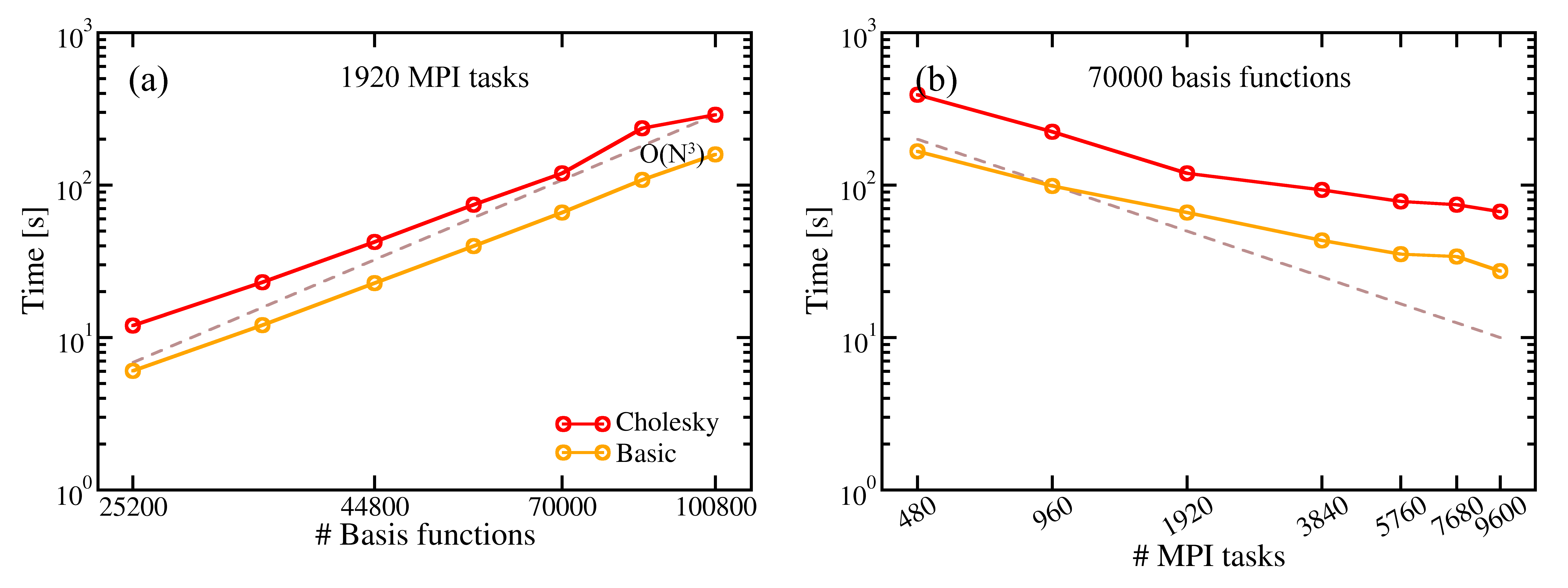}
\caption{Scaling of the computation of the density matrix using orbital minimization method, with respect to (a) the number of basis functions and (b) the number of MPI tasks. The number of MPI tasks in (a) is 1,920. The number of basis functions in (b) is 70,000. Shown here is the ideal case of OMM, where the CG line search of the OMM energy functional minimum requires only one step to converge. In practical SCF calculations, the number of line searches in OMM can only be reduced to one after several SCF steps. ``Basic'' refers to the method that directly handles the generalized eigenproblem. ``Cholesky'' refers to the method that applies Cholesky factorization to transform the generalized problem to the standard form before minimization. Ideal scaling is indicated by the dashed line.}
\label{fig:omm_flavors}
\end{figure*}

Compared in the second and third rows of Table \ref{tab:omm_cg} is another factor that has an impact on the number of CG line searches in libOMM, i.e., the method used to minimize the OMM functional. The basic algorithm directly follows the recipe in Eq. \ref{eq:omm_energy}, but Eq. \ref{eq:omm_energy} can also be minimized by first transforming the generalized eigenproblem to a standard problem based on Cholesky factorization. As shown in Table \ref{tab:omm_cg}, minimizing the OMM functional in the context of a standard eigenproblem (Cholesky, the third row in the table) contributes to a decrease in the number of line searches. This acceleration of the CG line search, however, comes at the price of the additional complexity required by the eigenproblem transformation. Fig. \ref{fig:omm_flavors} shows the comparison of the computational time of one CG line search in libOMM with the basic flavor versus the Cholesky flavor. The two flavors scale similarly, with respect to both the number of basis functions (Fig. \ref{fig:omm_flavors} (a), from 25,200 to 100,800 basis functions) and the number of MPI tasks (Fig. \ref{fig:omm_flavors} (b), from 480 to 9,600 MPI tasks). The Cholesky flavor is consistently slower than the basic flavor by a factor of 2 $\sim$ 4, due to the eigenproblem transformation and the corresponding back-transformation of Wannier function coefficients. Also reflected in Fig. \ref{fig:omm_flavors} is the shortest time to compute the density matrix using OMM, which is the basic flavor that converges in one CG line search. Indicated by Table \ref{tab:omm_cg} and Fig. \ref{fig:omm_flavors}, the most promising approach that could be used in practical calculations is the combination of a few ELPA steps followed by the basic flavor of OMM, whose convergence is guaranteed within one CG iteration. To further improve the performance of this solver, future work will include the inclusion in the ELSI interface of a preconditioned libOMM flavor, which has already proven to efficiently speed up the line search convergence \cite{libomm_corsetti_2014,precondition_lu_2017}; a spectral slicing method to separately evaluate the eigenstates near the Fermi level and thus to enable the proper handling of fractional occupation numbers; the sparse linear algebra via routines implemented in the PSPBLAS (Parallel SParse BLAS) library \cite{psp_yang_2017}; and ultimately the extension of OMM to a linear scaling solver as originally proposed \cite{omm_mauri_1993,omm_ordejon_1993,omm_mauri_1994,omm_ordejon_1995}.

\subsection{PEXSI}
\label{subsec:results_pexsi}
\begin{figure*}[h!]
\centering
\includegraphics[width=\textwidth]{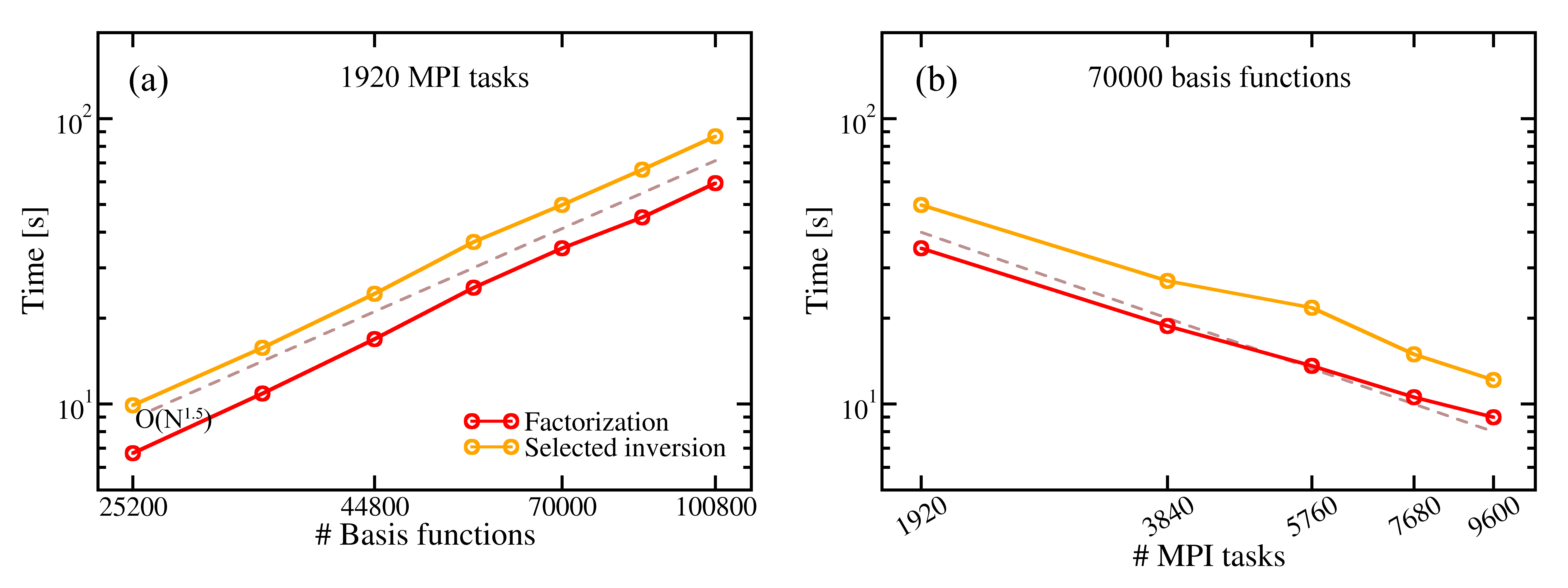}
\caption{Scaling of the two key computational steps of the PEXSI DFT driver, namely the numerical factorization and the selected inversion, with respect to (a) the number of basis functions and (b) the number of MPI tasks. The number of MPI tasks in (a) is 1,920. The number of basis functions in (b) is 70,000. Ideal scaling is indicated by the dashed lines. The 80 poles employed for the pole expansion in Eq. \ref{eq:pole} are independently evaluated in parallel. The numerical factorization and selected inversion of each pole are carried out using 1920/80 = 24 MPI tasks in (a), and \# MPI tasks/80 in (b).}
\label{fig:pexsi_steps}
\end{figure*}

\begin{figure*}[h!]
\centering
\includegraphics[width=\textwidth]{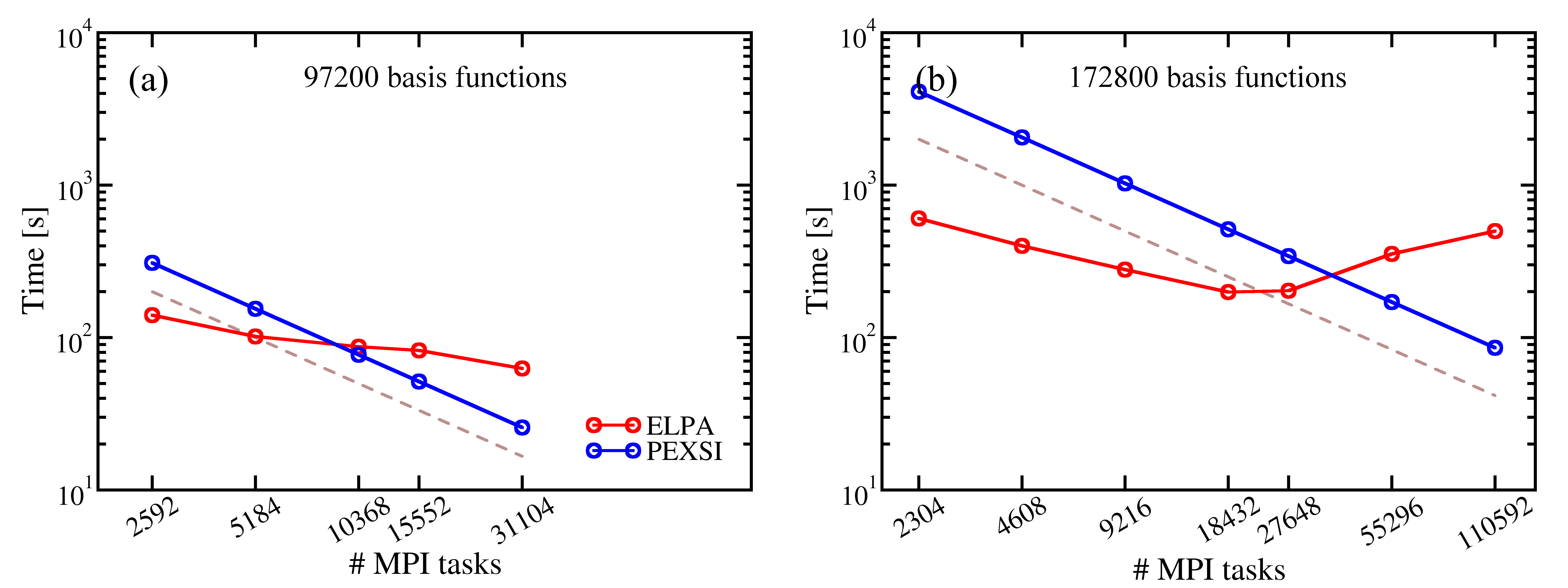}
\caption{Comparison of the strong scaling of the ELPA and PEXSI solvers. The number of basis functions is 97,200 in (a) and 172,800 in (b). The matrices are from DGDFT code. There are 48 poles employed in the PEXSI pole expansion. Ideal scaling is indicated by the dashed lines.}
\label{fig:pexsi_scaling}
\end{figure*}

As noted in Section \ref{subsec:pexsi_intro}, PEXSI exploits two levels of parallelization: the first level is the parallel evaluation of each pole in the pole expansion (Eq. \ref{eq:pole}), and the second level is the parallel numerical factorization and selected inversion at each pole. MPI tasks are divided into several groups with one pole assigned to each group. Fig. \ref{fig:pexsi_steps} (a) shows that both steps scale as $O(N^{1.5})$ for the graphene model, which is in agreement with the theoretical prediction for quasi-2D systems. The selected inversion step is slightly more expensive than the numerical factorization step. As shown in the strong scaling plot in Fig. \ref{fig:pexsi_steps} (b), both the numerical factorization and the selected inversion scale almost ideally to at least 9,600 MPI tasks. The number of MPI tasks shown in Fig. \ref{fig:pexsi_steps} (b) should be divided by the number of poles, 80, to reflect the scaling of numerical factorization and selected inversion at each pole. Since PEXSI has been shown to scale to several thousands of MPI tasks \cite{pselinv_jacquelin_2016}, the performance reported in Fig. \ref{fig:pexsi_steps} (b), which measures scalability up to 120 tasks per pole, is still far from the scalability limit. To further demonstrate the strong scaling of the PEXSI solver, Fig. \ref{fig:pexsi_scaling} shows the wall clock time used by PEXSI for a graphene model consisting of 6,480 atoms (97,200 basis functions) using 2,592 to 31,104 MPI tasks (Fig. \ref{fig:pexsi_scaling} (a)) and a graphene model consisting of 11,520 atoms (172,800 basis functions) using 2,304 to 110,592 MPI tasks (Fig. \ref{fig:pexsi_scaling} (b)). These tests are performed using the ELSI interface as implemented in the DGDFT software package. The ELPA eigensolver is also included as a reference. For both models, PEXSI exhibits nearly ideal strong scaling and eventually outperforms ELPA as the number of MPI tasks becomes sufficiently large. The ELPA solver ceases to scale beyond 18,432 MPI tasks for the 172,800-atom system.

\section{Conclusions}
\label{sec:conclusion}
Materials simulations based on Kohn-Sham density-functional theory require solving an eigenvalue problem repeatedly in an iterative procedure designed to obtain the ground state electron density of a poly-atomic system. Although this is a well studied subject in numerical linear algebra, it constitutes the bottleneck in large-scale calculations. A number of new approaches have emerged in the last few years. These approaches have different features and performance characteristics. Proper use of these approaches requires a good understanding of the pros and cons of each approach, and the input and output of specific algorithms. ELSI is designed to provide a common interface that allows users to easily choose an appropriate solver. Although the choice of the best solver often depends on a number of factors such as the problem size and the available computational resource, the benchmark results presented in this paper provide some general guidance on how to make these choices. In particular, we have shown different regimes in which one approach outperforms others and the crossover points between these regimes.

Finally, we demonstrated how different solvers can be organized in a common framework to enable easy integration with a vast number of electronic structure software packages. We anticipate that the number of new approaches to solving eigenvalue problems related to KS-DFT will continue to increase. We hope that ELSI will become a focal point for the community to integrate, comparatively assess and, ultimately, adopt this diverse ecosystem in a simple, effective fashion.

\section{Acknowledgments}
\label{sec:acknowledgments}
This work was supported by the National Science Foundation under grant number 1450280. We thank the National Energy Research Scientific Computing Center (NERSC) for the computational resources. This work also used resources of the Argonne Leadership Computing Facility, which is a DOE Office of Science User Facility supported under contract DE-AC02-06CH11357. Regarding the establishment and improvement of the ELSI API, we especially appreciate the fruitful discussions with and the feedback from many fellow researchers in the electronic structure community, including developers of the BigDFT, CP2K, DGDFT, FHI-aims, SIESTA code projects and of CECAM's Electronic Structure Library (http://esl.cecam.org). AG thanks EU H2020 grant 676598 (`MaX: Materials at the eXascale' CoE), Spain's MINECO (FIS2012-37549-C05-05, FIS2015-64886-C5-4-P and the `Severo Ochoa' program grant SEV-2015-0496), and GenCat (2014 SGR 301). The work of JL was partially supported by the National Science Foundation under grant number DMS-1127914 to the Statistical and Applied Mathematical Sciences Institute. VB particularly acknowledges the experiences shared by many of the co-authors of the FHI-aims code over many years, for instance, regarding parallelization strategies or the handling of ill-conditioned overlap matrices, from which we learned during the development of the ELSI interface - especially Dr. Ville Havu (Aalto University) and Dr. Rainer Johanni (Munich; now deceased).

\appendix
\section{Technical Settings in FHI-aims Calculations}
\label{app:fhiaims}
\begin{figure*}[h!]
\centering
\includegraphics[width=\textwidth]{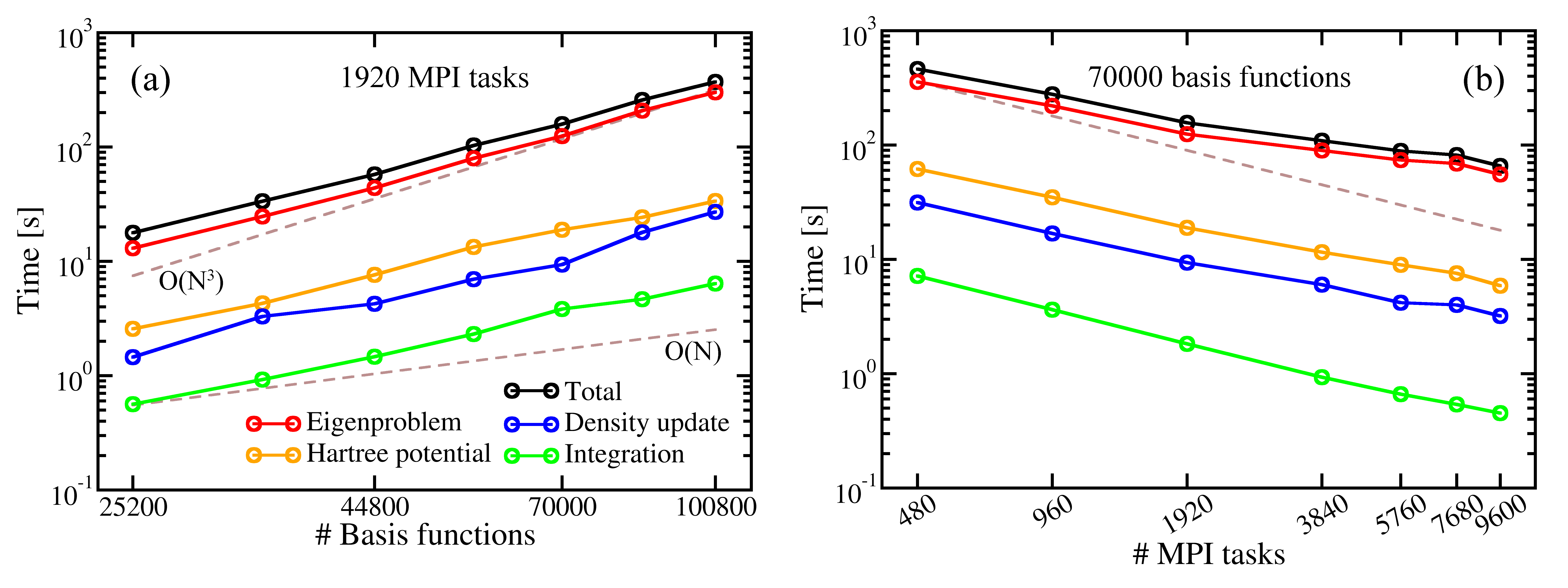}
\caption{Scaling of the key computational steps of the DFT-PBE calculations in FHI-aims, with respect to (a) the number of basis functions and (b) the number of MPI tasks. The number of MPI tasks in (a) is 1,920. The number of basis functions in (b) is 70,000. Ideal scaling is indicated by the dashed lines. The key steps are the evaluation of the Hartree potential, the numerical integrations of the Hamiltonian matrix elements, the update of the electron density and its gradient, and solving the Kohn-Sham eigenproblem using the ELPA eigensolver library. Shown here are timings corresponding to one SCF iteration, not accumulated timings in a complete SCF cycle.}
\label{fig:fhiaims}
\end{figure*}

The benchmark calculations reported in Figs. \ref{fig:sovlers}, \ref{fig:redist}, \ref{fig:init} \ref{fig:elpa_steps}, \ref{fig:omm_flavors}, and \ref{fig:pexsi_steps} in Section \ref{sec:results} are KS-DFT calculations performed with the FHI-aims code \cite{fhiaims_blum_2009,nao_havu_2009}, PBE exchange-correlation functional \cite{pbe_perdew_1996}, ``tier1'' numeric atom-centered orbital (NAO) basis set (see Table 1 in Ref. \cite{fhiaims_blum_2009}, ``light'' numerical settings, and a $1 \times 1 \times 1$ k-grid ($\Gamma$ point). In order to place the timings reported in Fig. \ref{fig:sovlers} into perspective with respect to the other parts of a KS-DFT calculation, Fig. \ref{fig:fhiaims} shows timings for all other important computational steps in the corresponding FHI-aims calculations, obtained on the same hardware and in the same runs as the results shown in Fig. \ref{fig:sovlers}. The main additional steps are executed on a real-space grid and include the Hartree potential evaluation, the numerical integrations of the Hamiltonian matrix elements, and the update of the electron density and its gradients, all implemented in a near $O(N)$ fashion and efficiently parallelized in FHI-aims. Refs. \cite{fhiaims_blum_2009,nao_havu_2009} provide a more detailed account of the algorithms involved.

\section{Technical Settings in DGDFT Calculations}
\label{app:dgdft}

The benchmark calculations reported in Figs. \ref{fig:pexsi_scaling} in Section \ref{sec:results} are KS-DFT calculations performed with DGDFT \cite{alb_lin_2012,dgdft_hu_2015} using the PBE exchange-correlation functional \cite{pbe_perdew_1996}. The global system is partitioned into $36\times 36$ and $48\times 48$ elements for the system containing 6,480 and 11,520 atoms, respectively. The number of adaptive local basis functions (ALB) per atom is 15, which is sufficient for the error of the total energy per atom and the maximum error of the force to be below $10^{-3}$ Hartree and $10^{-3}$ Hartree/Bohr, respectively. The DG penalty parameter is chosen to be 5.0, and the kinetic energy cutoff to generate the ALBs is set to 40 Hartree. The number of poles used by PEXSI is 48.

\bibliographystyle{elsarticle-num}
\bibliography{vwzyu}

\end{document}